\documentclass[secnumarabic,aps,prd,floats,floatfix,nofootinbib,showkeys,showpacs,10pt]{revtex4-1}

\usepackage[bookmarks=false]{hyperref}
\usepackage{amsfonts}
\usepackage{amsmath}
\usepackage{amssymb}
\usepackage{natbib}
\usepackage{graphicx}
\usepackage{enumitem}

\def\apj{ApJ}
\def\apjl{ApJL}

\def\jcap{JCAP}
\def\mnras{MNRAS}
\def\physrep{Physics Reports}
\def\prd{Phys. Rev. D}
\def\apjs{ApJS}

\graphicspath{{./fig/}}
\begin{document}
\title{Cosmological perturbation theory in Generalized Einstein-Aether models}
\author{Richard~A.~\surname{Battye}}
\email{richard.battye@manchester.ac.uk}
\author{Francesco~\surname{Pace}}
\email{francesco.pace@manchester.ac.uk}
\author{Damien~\surname{Trinh}}
\email[Corresponding author: ]{damien.trinh@postgrad.manchester.ac.uk}

\affiliation{Jodrell Bank Centre for Astrophysics, School of Physics and Astronomy, The University of Manchester, 
Manchester, M13 9PL, U.K.}

\label{firstpage}

\date{\today}

\begin{abstract}
We investigate the evolution of cosmological perturbations in models of dark energy described by a time-like unit normalized vector field specified by a general function $\mathcal{F}(\mathcal{K})$, so-called Generalized Einstein-Aether models. First we study the background dynamics of such models via a designer approach in an attempt to model this theory as dark energy. We find that only one specific form of this designer approach matches $\Lambda \mathrm{CDM}$ at background order and we also obtain a differential equation which $\mathcal{F}(\mathcal{K})$ must satisfy for general $w$CDM cosmologies. We also present the equations of state for perturbations in Generalized Einstein-Aether models, which completely parametrize these models at the level of linear perturbations. A generic feature of modified gravity models is that they introduce new degrees of freedom. By fully eliminating these we are able to express the gauge invariant entropy perturbation and the scalar, vector, and tensor anisotropic stresses in terms of the perturbed fluid variables and metric perturbations only. These can then be used to study the evolution of perturbations in the scalar, vector, and tensor sectors and we use these to evolve the Newtonian gravitational potentials. 
\end{abstract}
\setlength{\abovedisplayskip}{12pt plus 2pt minus 9pt}
\setlength{\belowdisplayskip}{12pt plus 2pt minus 9pt}
%\pacs{98.80.-k, 95.36.+x}

%\keywords{Cosmology; vector field; Einstein-Aether; dark energy; equation of state}

\maketitle
\section{Introduction}
The nature of dark energy remains one of the biggest unsolved problems in cosmology. Numerous models of dark energy and modified gravity theories have been constructed \cite{modgrav} in an attempt to describe cosmological observations \cite{Perlmutter,Riess,Spergel}, with varying degrees of success. Perhaps the simplest and most successful of these is the cosmological constant which is remarkably consistent with recent observations \cite{PlanckCP,PlanckDE}. However, other models must be studied in case they provide a more suitable description or otherwise to rule them out all together, both theoretically and observationally. With the advent of surveys such as DES\footnote{\url{http://www.darkenergysurvey.org/}} \cite{DES}, Euclid\footnote{\url{http://www.euclid-ec.org/}} \cite{Euclid1,Euclid2,Euclid3}, LSST\footnote{\url{https://www.lsst.org/}} \cite{LSST1,LSST2}, and SKA\footnote{\url{https://www.skatelescope.org/}} \cite{SKA1,SKA2,SKA3,SKA4}, observational constraints on these models will undoubtedly become tighter.

An obvious way to modify gravity is to introduce a new field other than the metric and make dark energy a dynamical component. These models typically introduce scalar fields and many of these are encompassed by Horndeski \cite{Horndeski1,Horndeski2}, the most general scalar-tensor theory that gives rise to second-order equations of motion. This class of models include Quintessence \cite{QT1,QT2,QT3}, $k$-essence \cite{KE1,KE2}, Kinetic Gravity Braiding (KGB) \cite{KGB}, $f(R)$ gravity \cite{FRST1,FRST2,FRST3}, and many more. Indeed, it has already been shown that it is possible to achieve a dark energy fluid with $w=-1$ exactly in, for example, Quintessence and $k$-essence \cite{RBFP}, and for so-called `designer $f(R)$' \cite{DesignerFR}. However, there is no reason not to consider the new field to be a vector and indeed such vector-tensor theories have been shown to be able to give rise to a period of accelerated expansion even without potential terms \cite{Jiminez1,Jiminez2,Jiminez3,Jiminez4,Mattingly1,Mattingly2,Ferreira1,Ferreira2,Ferreira3}, and therefore provide an interesting avenue of research. In this paper we study so-called Einstein-Aether theories at background and perturbative order, where the vector field is constrained to be of time-like unit norm. First studied in \cite{Mattingly1}, it was shown that the model would in fact slow the expansion of the universe \cite{Carroll}. However, more recently, modifications to this theory have been shown to allow it to describe dark energy and still be compatible with observations \cite{Ferreira1,Ferreira2,Ferreira3}. This is done by introducing non-canonical kinetic terms parametrized by a free function $\mathcal{F}(\mathcal{K})$, where $\mathcal{K}$ determines the kinetic terms for the vector field. In principle this could take on any functional form and in previous work in this area specific forms were chosen to work with. However, as with designer $f(R)$, we will choose a background evolution of the universe and allow that to dictate the form of $\mathcal{F}(\mathcal{K})$ in a `designer $\mathcal{F}(\mathcal{K})$' model.

At background order, despite the many complex models of dark energy all of these can be parametrized by specifying a single function of time, the equation of state parameter, $w_{\mathrm{de}}=P_{\mathrm{de}}/\rho_{\mathrm{de}}$. Exactly how $w_{\mathrm{de}}$ behaves as a function of time will of course depend on the theory, but at this order there is nothing else to measure which will tell us about the nature of dark energy, provided FRW spacetime symmetries are respected. At the level of linear perturbations various approaches have been developed in order to try to parametrize different theories \cite{EFT1,EFT2,EFT3,PPF1,PPF2,PPF3,PPF4,EOS1,EOS2,EOS3,GeneralSVT}. In this paper, we work with the Equation of State for perturbations (EoS) approach \cite{EOS1,EOS2,EOS3}. A generic feature of modified gravity models is that new degrees of freedom arise at the level of perturbations. The EoS approach packages the parametrization into the gauge invariant entropy perturbation, $\Gamma$, and anisotropic stress, $\Pi^{S}$, by eliminating these degrees of freedom in favour of the perturbed fluid variables and metric perturbations. The perturbed conservation equation, $\delta (\nabla_\mu T^\mu\hspace{0.1mm}_\nu)=0$, gives two evolution equations for the density perturbation, $\delta \rho$, and divergence of the velocity field, $\theta^S$. For example, in the synchronous gauge they are given by \begin{align} \label{ConsEq1}
\left(\frac{\delta}{1+w} \right)' &= -k^2 \theta^S - \frac{1}{2}h' - \frac{3\mathcal{H}}{1+w}w\Gamma, \\\nonumber\\ \label{ConsEq2}
(1+w)\theta^{S'} = \mathcal{H}(1+w)&\left( 3\frac{dP}{d\rho}-1 \right)\theta^S + \frac{dP}{d\rho}\delta + w\Gamma +\frac{2}{3}w \Pi^S,
\end{align} where primes denote conformal time differentiation and $\mathcal{H}$ is the conformal Hubble parameter. The metric perturbations, $h$ and $\eta$, are evolved via Einstein's equation. However, the forms of $\Pi^S$ and $\Gamma$ are not known and hence \eqref{ConsEq1} and \eqref{ConsEq2} are not closed. If we can somehow specify $\Gamma$ and $\Pi^{S}$ as linear functions of the perturbed fluid variables, metric perturbations, and their derivatives only, these equations close, i.e. we wish to write $\Gamma = \Gamma(\delta, \theta^S, h', \eta, ...)$ and $\Pi^S = \Pi^S(\delta, \theta^S, h', \eta, ...)$, or equivalently in terms of the dark energy ($\mathrm{de}$) and matter ($\mathrm{m}$) fluid variables, $\Gamma = \Gamma\left( \delta_{\mathrm{de}}, \theta^S_{\mathrm{de}}, \delta_{\mathrm{m}}, \theta^S_{\mathrm{m}}\right) $ and $\Pi^S = \Pi^S\left( \delta_{\mathrm{de}}, \theta^S_{\mathrm{de}}, \delta_{\mathrm{m}}, \theta^S_{\mathrm{m}}\right) $. Our approach is to eliminate the internal degrees of freedom describing the dynamics of the modified gravity theory, via expressions for $\delta$ and $\theta^S$, supplemented by the equation of motion for the vector field. In principle, the equations of motion and hence the perturbed fluid variables have already been derived in \cite{Ferreira1,Ferreira2}, for example, although the equations of state have not been computed. However, in most of the previous work the so-called `acceleration' term has not been included, corresponding to the $c_4$ term in \cite{Mattingly2}. This term is often either completely ignored or argued that a transformation of the coefficients can remove it. However, we discuss later why this isn't true in general and so keep the $c_4$ term in our subsequent analysis. In particular, we extend on previous work done by including the $c_4$ term for $\mathcal{F}(\mathcal{K})$ theories in so-called Generalized Einstein-Aether, as well as using the EoS formalism.

Although in this paper we use a specific Lagrangian to work with, one of the advantages of the EoS approach is that it allows the computation of cosmological perturbations in a model independent way. In \cite{EOS3} this approach was applied to generic scalar-tensor theories by specifying only the field content of the Lagrangian and nothing specific about its functional form. This approach also provides a set of modifications that are, in principle, easy to insert into numerical codes. Equations of state have already calculated for various different classes of theories, for example, the elastic dark energy (EDE) \cite{EDE}, which was shown to be equivalent to Lorentz violating, massive gravity theories \cite{MassGrav}. They have also been calculated for general scalar-tensor theories \cite{EOS3} and in particular Quintessence, $k$-essence, KGB, and Horndeski theories \cite{Horndeski2}. In these cases, the degree of freedom to be eliminated is related to the perturbed scalar field, $\delta \varphi$, and its derivatives. This was also shown to be the case for $f(R)$ gravity and was studied in \cite{fRDE}. In this paper we apply the EoS approach to Generalized Einstein-Aether theories. The expressions for $\Gamma$ and $\Pi^S$ are shown in \autoref{table:GP} for some of these theories, in the synchronous gauge, where $\left\lbrace \mathcal{A}_i\right\rbrace $ are functions of background quantities and $c_{\mathrm{s}}^2=\delta P/\delta \rho$ is the squared sound speed of scalar perturbations. We do not provide the expressions in $f(R)$ gravity here as they are quite complicated, however they are presented in \cite{fRDE}.
\begin{table}
\begin{center}
	{\renewcommand{\arraystretch}{1.6}\begin{tabular}{ c | c | c }
		\textbf{Theory} &\textbf{Scalar anisotropic stress}, $w\Pi^S$ & \textbf{Entropy perturbation}, $w\Gamma$ \\
		\hline
		Minimally coupled scalar fields & $0$ & $\left(c_{\mathrm{s}}^2 - \frac{dP}{d\rho} \right)\left[\delta + 3H(1+w)\theta^S \right]   $\\ 
		KGB & $0$ &$\mathcal{A}_1\delta +\mathcal{A}_2\theta^S + \mathcal{A}_3h'+\mathcal{A}_4h''$ \\
		EDE & $\frac{3}{2}\left(w-c_{\mathrm{s}}^2 \right)\left[\delta- 3(1+w)\eta \right]  $ & $0$ \\
	\end{tabular}}
\end{center}
\caption{Expressions for $\Gamma$ and $\Pi^S$ in terms of the dark energy perturbed fluid variables and metric perturbations for some dark energy models and modified gravity theories, in the synchronous gauge.} \label{table:GP}
\end{table}

This paper is organized as follows. In \autoref{sect:GEA} we present the model for Generalized Einstein-Aether and derive the equations of motion. We also briefly mention sub-cases to this model that have been studied previously. We then study the theory at linear perturbative order (\autoref{sect:Pert}) in the scalar, vector, and tensor sector and present expressions for the perturbed fluid variables in both the conformal Newtonian and synchronous gauges. We then proceed to derive the gauge invariant equations of state for perturbations (\autoref{sect:EoS}) by eliminating all the internal degrees of freedom that arise from introducing the vector field. From these we also study the evolution of the Newtonian gravitational potentials. We then conclude in \autoref{sect:Conc} and discuss future steps.

Natural units are used throughout with $c = \hbar = 1$ and the metric signature is $(-,+,+,+)$.

\section{Generalized Einstein-Aether Field Equations} \label{sect:GEA}
\subsection{Field equations}
The Lagrangian for Generalized Einstein-Aether is \cite{Ferreira1} \begin{equation} \label{General Lagrangian}
16 \pi G\mathcal{L}_A = M^2 \mathcal{F}(\mathcal{K}) + \lambda(g_{\mu \nu}A^\mu A^\nu +1),
\end{equation} where we introduce the vector field $A^\mu$, which is known as the Aether field. The scalar $\mathcal{K}$ is defined by \begin{equation} \mathcal{K} = \frac{1}{M^2}K^{\alpha \beta}\hspace{0.1mm}_{\mu \nu} \nabla_\alpha A^\mu \nabla_\beta A^\nu \end{equation} and the rank-4 tensor is defined by \begin{equation} \label{Kinetic Tensor}
K^{\alpha \beta} \hspace{0.1mm} _{\mu \nu} = c_1 g^{\alpha \beta} g_{\mu \nu} + c_2\delta^\alpha _\mu \delta^\beta _\nu + c_3\delta^\alpha _\nu \delta^\beta _\mu + c_4 A^\alpha A^\beta g_{\mu \nu}.
\end{equation} Here, $\left\lbrace c_i\right\rbrace $ are dimensionless constants and $M$ has dimensions of mass. The `kinetic tensor', $K^{\alpha \beta} \hspace{0.1mm} _{\mu \nu}$, determines the derivative squared terms of the Aether field. Similar to generalization of Quintessence to $k$-essence, the kinetic terms have been modified to an arbitrary, dimensionless function $\mathcal{F}(\mathcal{K})$. An important feature of Einstein-Aether models is the presence of the Lagrange multiplier $\lambda$. This will constrain the Aether field to have a time-like unit norm. As we will see, this will also have an effect on the propagating degrees of freedom at the perturbative level. 

The full action that we will study is then \begin{equation} \label{action}
S = \int d^4 x \sqrt{-g} \left( \frac{1}{16 \pi G}R + \mathcal{L}_A\right)  +S_{\mathrm{m}},
\end{equation} where the action for the matter fields, $S_\mathrm{m}$, does not couple directly to the Aether field. The equations of motion can now be obtained by varying \eqref{action} with respect to each degree of freedom i.e. $\lambda$, $A^\mu$, and $g^{\mu \nu}$. Variation with respect to $\lambda$ yields the constraint $g_{\mu \nu}A^\mu A^\nu = -1$. The equation of motion for the Aether field, $A^\mu$, is \begin{equation} \label{Vector EoM}
\nabla_\alpha (\mathcal{F}_\mathcal{K}J^{\alpha}\hspace{0.1mm}_{\mu}) - c_4 \mathcal{F}_{\mathcal{K}}A^\alpha \nabla_\alpha A^\nu \nabla_\mu A_\nu = \lambda A_\mu,
\end{equation} where we define $J^{\alpha}\hspace{0.1mm}_\mu = K^{\alpha \beta} \hspace{0.1mm} _{\mu \nu} \nabla_\beta A^\nu $ and $\mathcal{F}_{\mathcal{K}} = \frac{d\mathcal{F}}{d\mathcal{K}}$, and variation with respect to the metric gives Einstein's equation in the form \begin{equation} \label{Einstein Equation}
G_{\mu \nu} = 8 \pi G T_{\mu \nu} + U_{\mu \nu},
\end{equation} where $T_{\mu \nu}$ is the energy-momentum tensor for the matter fields only. All contributions from the Aether field are included in $U_{\mu \nu}$  which takes the form \begin{align} 
U_{\alpha \beta} =& \hspace{1mm} \nabla_\mu \left[ \mathcal{F}_{\mathcal{K}} \left( J_{(\alpha}\hspace{0.1mm}^{\mu} A_{\beta )} - J^\mu \hspace{0.1mm} _{(\alpha}A_{\beta)} - J_{(\alpha \beta)}A^\mu \right) \right] +\lambda A_\alpha A_\beta +\frac{1}{2} M^2 \mathcal{F} g_{\alpha \beta} \nonumber
\\&+ c_1 \mathcal{F}_{\mathcal{K}} \left(\nabla_\mu A_\alpha \nabla^\mu A_\beta - \nabla_\alpha A_\mu \nabla_\beta A^\mu\right) +c_4 \mathcal{F}_\mathcal{K}A^\mu A^\nu \nabla_\mu A_\alpha \nabla_\nu A_\beta,
\end{align} where brackets around indices denote symmetrization, i.e. $J_{\left( \alpha \beta\right)} = \frac{1}{2} \left(J_{\alpha \beta} + J_{\beta \alpha} \right)  $.

Using \eqref{Vector EoM} to eliminate $\lambda$, we find that	
\begin{align} \label{E-A Tensor c_4}
U_{\alpha \beta} =& \hspace{1mm} \nabla_\mu \left( \mathcal{F}_{\mathcal{K}} \left[ J_{(\alpha}\hspace{0.1mm}^{\mu} A_{\beta )} - J^\mu \hspace{0.1mm} _{(\alpha}A_{\beta)} - J_{(\alpha \beta)}A^\mu \right] \right) \nonumber
\\&+ c_1 \mathcal{F}_{\mathcal{K}} \left(\nabla_\mu A_\alpha \nabla^\mu A_\beta - \nabla_\alpha A_\mu \nabla_\beta A^\mu\right) +c_4 \mathcal{F}_\mathcal{K}A^\mu A^\nu \nabla_\mu A_\alpha \nabla_\nu A_\beta \nonumber
\\& +\left[c_4 \mathcal{F}_{\mathcal{K}}A^\mu A^\nu \nabla_\mu A^\tau \nabla_\nu A_\tau - A^\nu \nabla_\mu (\mathcal{F}_{\mathcal{K}} J^{ \mu}\hspace{0.1mm}_\nu ) \right] A_\alpha A_\beta +\frac{1}{2} M^2 \mathcal{F} g_{\alpha \beta}.
\end{align} The first line arises due to the metric variation in the Christoffel symbols \cite{Carroll,Halle}, the second line comes from the variation in the $c_1$ and $c_4$ terms of \eqref{Kinetic Tensor}, and the third line is due to the variation of the Lagrange multiplier and $\sqrt{-g}$ terms.

\subsection{Background dynamics} \label{sect:Background}

We will assume a background cosmology described by the FRW metric, \begin{equation}
ds^2 = -dt^2 + a(t)^2\delta_{ij} dx^i dx^j,
\end{equation} and that $A^\mu = (1,0,0,0)$. The reason for this choice of $A^\mu$ is to satisfy the unit norm constraint and to be compatible with the symmetries of FRW. Taking $U_{\mu \nu}$ to be the energy momentum tensor of a perfect fluid, then from $U_{00}$ and $U_{ij}$ we find that the background energy density and pressure are \begin{align} \label{rho}
&\rho_A = 3 \alpha H^2 \left( \mathcal{F}_{\mathcal{K}} - \frac{\mathcal{F}}{2\mathcal{K}}\right), \\
 \label{Pressure}
P_A= \alpha & \left[ 3H^2\left(\frac{\mathcal{F}}{2\mathcal{K}}-\mathcal{F}_{\mathcal{K}}  \right) - \dot{\mathcal{F}_{\mathcal{K}}} H -\mathcal{F}_{\mathcal{K}}\dot{H}\right],
\end{align} where $\alpha = c_1 + 3c_2 + c_3$, over-dots denote differentiation with respect to cosmic time, $t$, and \begin{equation} \label{Scalar K}
\mathcal{K} = \dfrac{3\alpha H^2}{M^2}.
\end{equation}
Note that we have absorbed a factor of $8\pi G$ into $U_{\mu\nu}$. We can also check that $P_A$ and $\rho_A$ satisfy the energy conservation equation \begin{equation}\label{Background ConsEq}
\dot{\rho}_A = -3H(\rho_A+ P_A),
\end{equation} as they should by construction of \eqref{Einstein Equation}. Note that the $c_4$ term plays no role in the background dynamics. 

The time-time component of Einstein's equation gives the modified Friedmann equation as \begin{equation} \label{Modified Friedmann}
(1- \alpha \mathcal{F}_{\mathcal{K}})H^2 +\frac{1}{6}\mathcal{F}M^2= \frac{8 \pi G}{3} \rho_m.
\end{equation} If we were to demand that the theory is indistinguishable from a cosmological constant at background order, then from \eqref{Modified Friedmann} we obtain the differential equation \begin{equation} \label{Friedmann FDE}
\mathcal{K} \frac{d \mathcal{F}}{d \mathcal{K}} - \frac{1}{2} \mathcal{F} = \frac{\Lambda}{M^2}, \end{equation} where we have substituted $H^2$ for $\mathcal{K}$ via \eqref{Scalar K}. The solution to this equation is  \begin{equation} \label{CCEA}
\mathcal{F} = B (\pm \mathcal{K})^{1/2} - \frac{2\Lambda}{M^2},
\end{equation} depending on the sign of $\mathcal{K}$ and where $B$ is an arbitrary integration constant. The case of a general power law has been studied in \cite{Ferreira1,Ferreira2,Ferreira3} as well as more exotic forms, for example see \cite{Halle,Meng}. Indeed, the functional form of $\mathcal{F}(\mathcal{K})$ must be specified at some point to make observational predictions. However, since $\mathcal{F}(\mathcal{K})$ could in principle be anything, it would be ideal if the form of $\mathcal{F}(\mathcal{K})$ could be found by specifying more standard parameters describing the background dynamics e.g. $w_{\mathrm{de}}$, $\Omega_{\mathrm{de},0}$, etc. Since any new dark energy model will at least have to be compatible with $\Lambda$CDM `globally', it makes sense to demand that Generalized Einstein-Aether must yield a $\Lambda$CDM cosmology and in turn, this will restrict the form of $\mathcal{F}(\mathcal{K})$. Since the background evolution of this model will be identical to $\Lambda$CDM, the effects of perturbations will become very important as it is only the dynamics at the perturbative level which will be able to distinguish this model from $\Lambda$CDM.

Let us now demand that the Aether field energy density and pressure obey a more general equation of state i.e. $P_A = w_{\mathrm{de}} \rho_A$, where $w_{\mathrm{de}}$ is constant. Since current observations do not yet sufficiently constrain anything other than constant $w_{\mathrm{de}}$ this is a reasonable assumption to make, however this may change in the near future. We can rewrite \eqref{Pressure} as \begin{equation} 
P_A = -\rho_A -\alpha (2\mathcal{K}\mathcal{F}_\mathcal{KK}+\mathcal{F}_\mathcal{K})\dot{H}
\end{equation} and so, \begin{equation} \label{DEwithHdot}
(1+w_{\mathrm{de}})M^2\left( \mathcal{KF}_{\mathcal{K}}-\frac{1}{2}\mathcal{F}\right) = - \alpha (2\mathcal{K}\mathcal{F}_\mathcal{KK}+\mathcal{F}_\mathcal{K})\dot{H},
\end{equation} where we have written $H^2$ in terms of $\mathcal{K}$. If we can write $\dot{H}=\dot{H}(\mathcal{K})$, then \eqref{DEwithHdot} will give us a differential equation to solve for $\mathcal{F}(\mathcal{K})$ satisfying a certain value of $w_{\mathrm{de}}$. 

We write the Friedmann equation as \begin{equation} \label{Friedmann Omega}
\left( \frac{H}{H_0} \right)^2 = \frac{\Omega_{\mathrm{m},0}}{a^3} + \frac{\Omega_{\mathrm{de},0}}{a^{3(1+w_{\mathrm{de}})}}, 
\end{equation} where we have defined $8\pi G \rho_{\mathrm{de}} = \rho_A$, $\textstyle\Omega_i=\frac{8\pi G}{3H^2}\rho_i$, and for this section only the subscript `$\mathrm{m}$' refers to matter with $P_{\mathrm{m}}=0$. Differentiating this and combining with $\eqref{Friedmann Omega}$ to eliminate $\Omega_{\mathrm{de},0}$ gives \begin{equation} \label{a^-3}
\frac{1}{a^3} = \frac{1}{w_{\mathrm{de}} \Omega_{\mathrm{m},0}}\left[ (1+w_{\mathrm{de}}) \left( \frac{H}{H_0} \right)^2 + \frac{2 \dot{H}}{3 {H_0}^2} \right].
\end{equation} We can also use the Raychaudhuri equation, given by \begin{equation}
\dot{H} + H^2 = -\frac{4\pi G}{3} \left[ \rho_\mathrm{m} + (1+3w_{\mathrm{de}})\rho_{\mathrm{de}} \right].
\end{equation} Inserting \eqref{rho} we have that \begin{equation}
\frac{\dot{H}}{{H_{0}}^2}+ \left( \frac{H}{H_0} \right)^2 = -\frac{\Omega_{\mathrm{m},0}}{2a^3} -\frac{M^2}{6{H_{0}}^2}(1+3w_{\mathrm{de}})\left( \mathcal{K}\mathcal{F}_{\mathcal{K}}-\frac{1}{2}\mathcal{F}\right),
\end{equation} and so using \eqref{a^-3} we find that \begin{equation}
\dot{H}(\mathcal{K}) = -\frac{M^2}{2}\left[ \frac{\mathcal{K}}{\alpha} + w_{\mathrm{de}} \left( \mathcal{K}\mathcal{F}_{\mathcal{K}}-\frac{1}{2}\mathcal{F}\right) \right].
\end{equation}

Therefore, the differential equation we must solve is then \begin{equation} \label{Background DE}
(1+w_{\mathrm{de}})\left( 2\mathcal{KF}_{\mathcal{K}}-\mathcal{F}\right) = (2\mathcal{K}\mathcal{F}_\mathcal{KK}+\mathcal{F}_\mathcal{K})\left[ \mathcal{K} + \frac{1}{2}\alpha w_{\mathrm{de}} \left( 2\mathcal{K}\mathcal{F}_{\mathcal{K}}-\mathcal{F}\right) \right].
\end{equation} For $w_{\mathrm{de}}= - 1$, then this reduces to \begin{equation}
(2\mathcal{K}\mathcal{F}_\mathcal{KK}+\mathcal{F}_\mathcal{K})\left[ \mathcal{K} - \frac{1}{2}\alpha  \left( 2\mathcal{K}\mathcal{F}_{\mathcal{K}}-\mathcal{F}\right) \right] = 0,
\end{equation}for which there are two branches of solutions,\begin{align} \label{Sol1}
\mathcal{F} &= \frac{2}{\alpha}\mathcal{K} + D (\pm\mathcal{K})^{1/2}, \\ \label{Sol2}
\mathcal{F} &= B (\pm\mathcal{K})^{1/2} + C, 
\end{align} again depending on the sign of $\mathcal{K}$ and where $B,C$ and $D$ are integration constants. If we insert \eqref{Sol1} into \eqref{Modified Friedmann} we find that the Friedmann equation becomes $\rho_{\mathrm{m}} = 0$ and therefore we ignore this branch of the solution. For the other branch, we see that \eqref{Sol2} is what we obtained before from demanding a cosmological constant, which sets $\textstyle C = - 2\Lambda/M^2= -6H_0^2\Omega_{\Lambda,0}/M^2$. Therefore, the only functional form for $\mathcal{F}$ which gives rise to an exact $\Lambda\mathrm{CDM}$ cosmology, at background order, is \eqref{CCEA}. More generally, we see that the initial conditions are related via \eqref{rho}, such that if we specify that today $\mathcal{F}(\mathcal{K}_0)= \mathcal{F}_0$, then it must be that \begin{equation} \label{Derivative IC}
\mathcal{F}_{\mathcal{K},0}= \frac{\Omega_{\rm{de},0}}{\alpha}+ \frac{\mathcal{F}_0}{2\mathcal{K}_0},
\end{equation} where $\mathcal{F}_{\mathcal{K},0} = \mathcal{F}_{\mathcal{K}}(\mathcal{K}_0)$ and $\mathcal{K}_0 = \mathcal{K}(a=1)$. Applying these initial conditions to \eqref{Sol2} we find that \begin{equation} \label{Analytical F}
\mathcal{F}= \left( \mathcal{F}_0 + \frac{6H_0^2\Omega_{\rm{de},0}}{M^2}\right) \left( \frac{\mathcal{K}}{\mathcal{K}_0}\right)^{1/2}-\frac{6H_0^2\Omega_{\rm{de},0}}{M^2}.
\end{equation}

\begin{figure}
	\centering
	\includegraphics[width=0.497\textwidth]{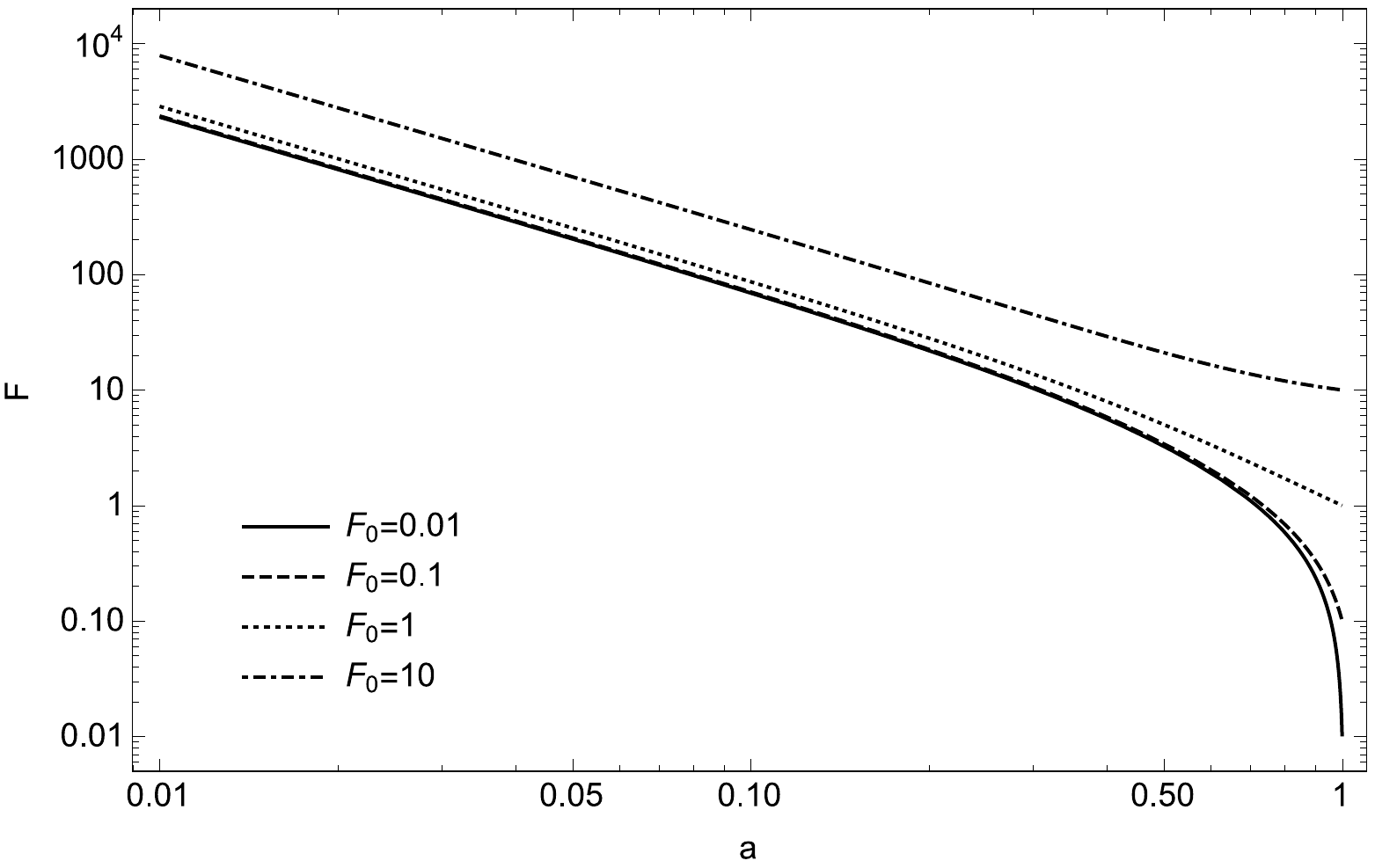}
	\includegraphics[width=0.497\textwidth]{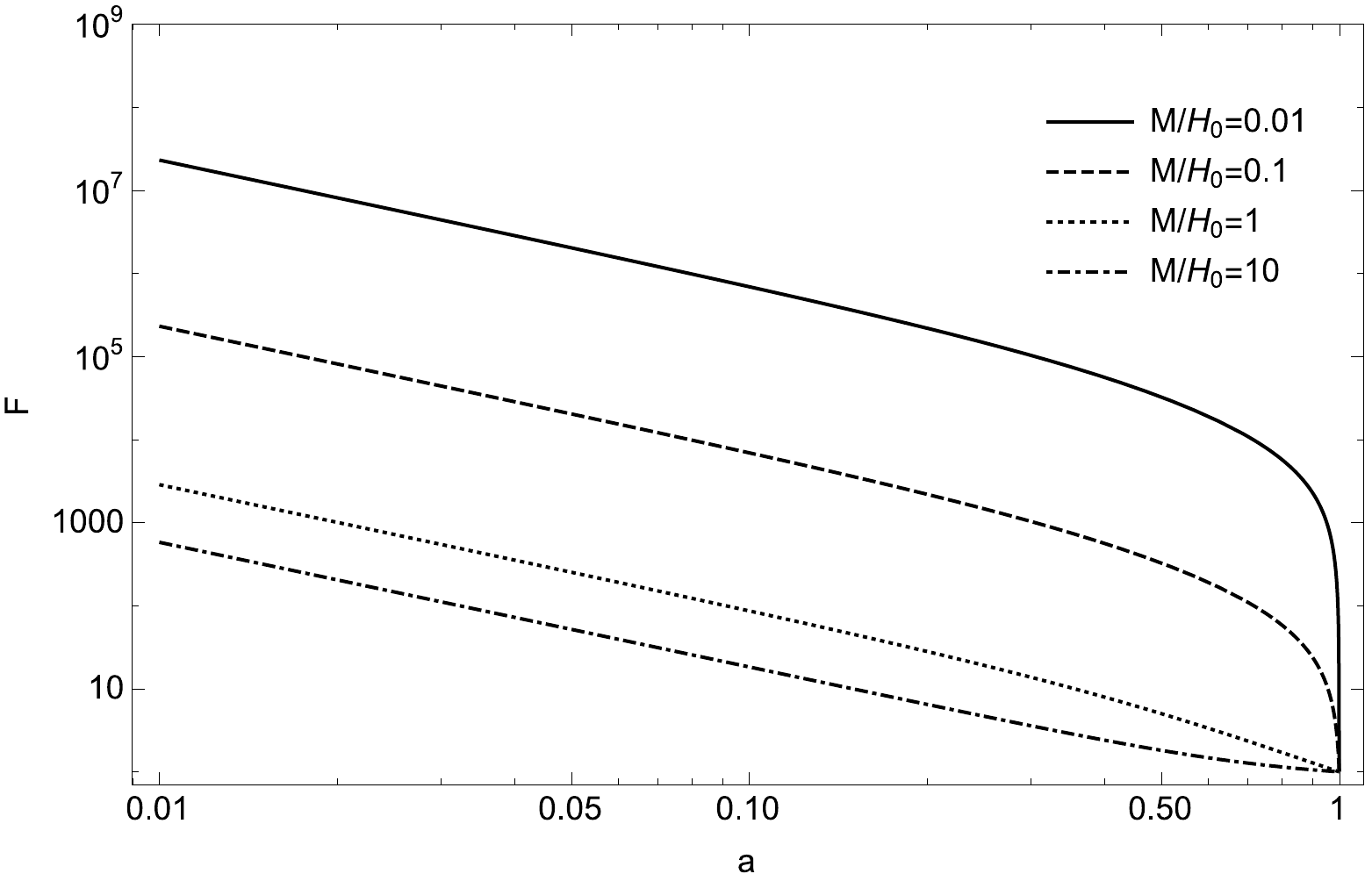}\\
	\includegraphics[width=0.497\textwidth]{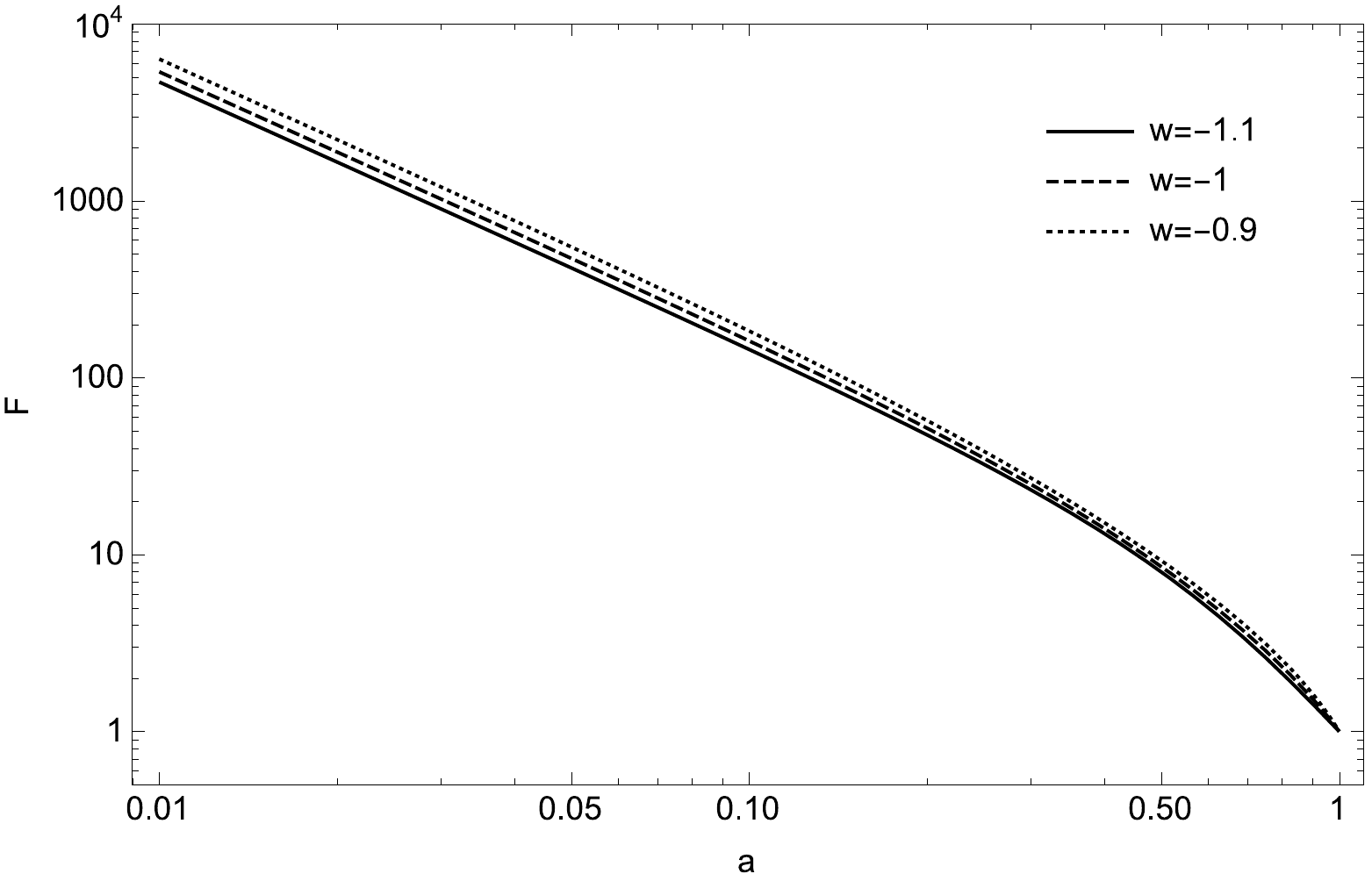}
	\includegraphics[width=0.497\textwidth]{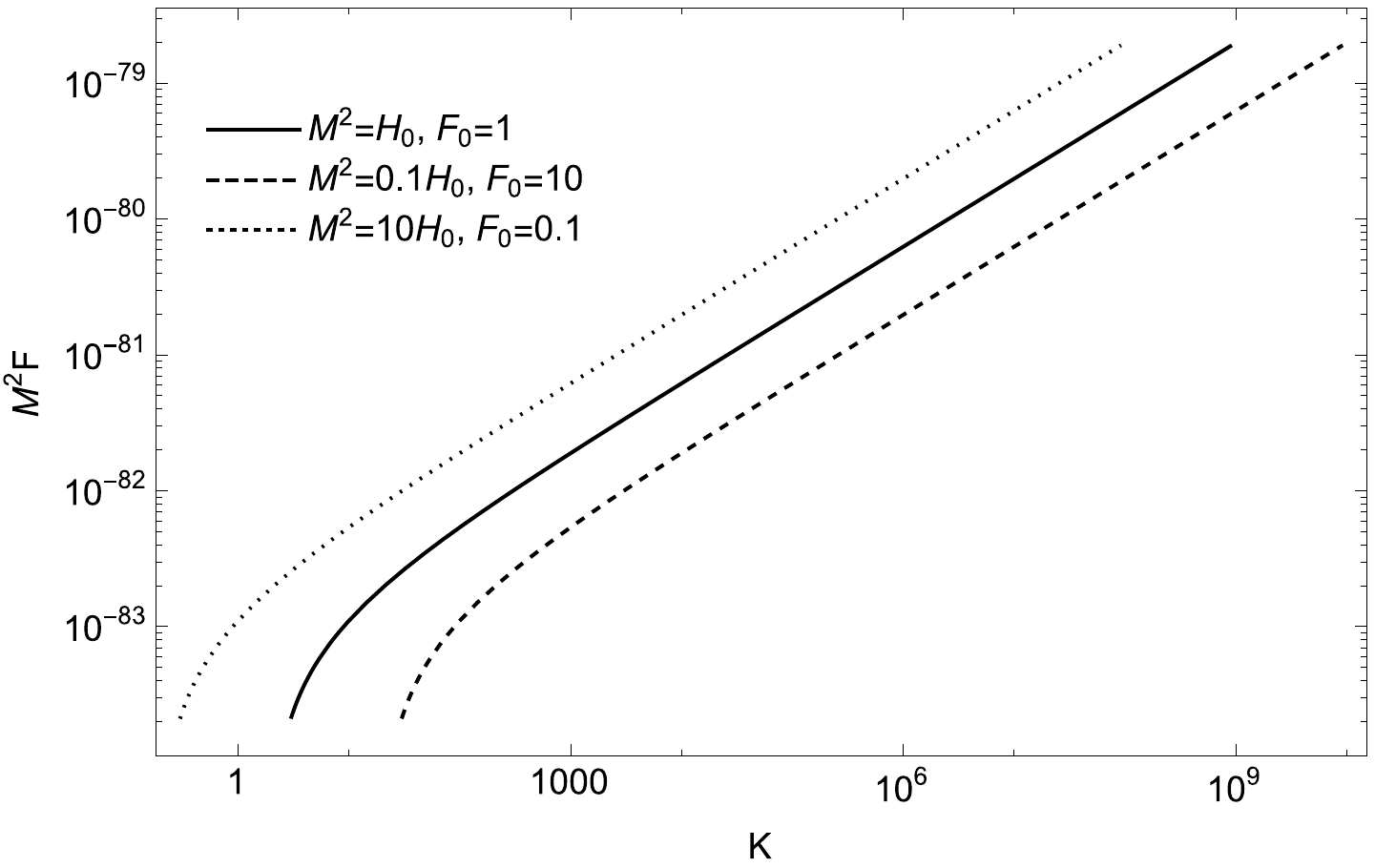}
	\caption{\textit{Top left panel}: Comparison of the evolution of $\mathcal{F}$ due to varying $\mathcal{F}_0$. In these models $M = H_0$ and $w_{\mathrm{de}}=-1$ are fixed.
		\textit{Top right panel}: Comparison of the evolution of $\mathcal{F}$ due to the variation of $M$, as a multiple of $H_0$. In these models $\mathcal{F}_0= 1$ and $w_{\mathrm{de}}=-1$ are fixed. \textit{Bottom left panel}: Comparison of the evolution of $\mathcal{F}$ for varying $w_{\mathrm{de}}$ close to $-1$. In these models $\mathcal{F}_0 = 1$ and $M=H_0$ are fixed. \textit{Bottom right panel}: Comparison of the evolution of $M^2\mathcal{F}$ for varying $M^2$ and $\mathcal{F}_0$, with $M^2\mathcal{F}_0/H_0^2 = 1$ and $w_{\mathrm{de}}=-1$ fixed. }
	\label{fig:F Evolution}
\end{figure}

At background order, we appear to have 5 parameters $\left\lbrace w_{\mathrm{de}}, \Omega_{\mathrm{de},0},\mathcal{F}_0,M, \alpha \right\rbrace$ which we must specify in order to compute $\mathcal{F}$. Varying $\alpha$ will vary the domain over which $\mathcal{F}$ varies as a function of $\mathcal{K}$. It may also seem that $\alpha$ will affect the functional form of $\mathcal{F}$, as it appears explicitly in \eqref{Background DE}. However, note that this is somewhat misleading because $\mathcal{K} \propto \alpha$ and the explicit dependence of $\alpha$ in \eqref{Background DE} is removed under a rescaling $\mathcal{K} \rightarrow \mathcal{K}/\alpha$. This can also be seen from \eqref{rho} and \eqref{Pressure}, where the factor of $\alpha$ is removed under the same rescaling. Therefore, $\alpha$ can take on any value for the purposes of the background evolution and so we will fix $\alpha = 1$ for the rest of this section.

The evolution of $\mathcal{F}$ for different $\left\lbrace w_{\mathrm{de}},\mathcal{F}_0, M \right\rbrace $ is shown in \autoref{fig:F Evolution}. We will fix $\Omega_{\mathrm{de},0}=0.691$ and $H_0 = 2.132\times10^{-42} h \, \mathrm{GeV}$, where $h=0.678$ \cite{PlanckCP}. To study the effect of varying $\mathcal{F}_0$ we will look to the analytical solution for $w_{\mathrm{de}}=-1$ in \eqref{Analytical F}, with $M=H_0$. The evolution of $\mathcal{F}$ will be such that it will be driven to $\mathcal{F}_0$ at $a=1$, as shown in \autoref{fig:F Evolution}. The parameter $\mathcal{F}_0$ is similar to designer $f(R)$ theories where the analogous parameter in \cite{fRDE} was called $B_0$. We see that in the past $\mathcal{F}$ is approximated well by a pure power law, corresponding to the behaviour of the first term in \eqref{Analytical F}, since this terms dominates in the past. For $\mathcal{F}_0 \gg 6H_0^2 \Omega_{\mathrm{de},0}/M^2$, this power law behaviour persists into the dark energy dominated era as $\mathcal{F}\rightarrow\mathcal{F}_0$. If $\mathcal{F}_0 \lesssim 6H_0^2 \Omega_{\mathrm{de},0}/M^2$ then for $\left( \mathcal{K}/\mathcal{K}_0\right)^{1/2} \gg 1$ the first term still dominates in \eqref{Analytical F} and we still observe the power law behaviour. However, as $\left( \mathcal{K}/\mathcal{K}_0\right)^{1/2} \rightarrow 1$ the second term in \eqref{Analytical F} becomes comparable to the first and so the power law behaviour is broken as $\mathcal{F}\rightarrow\mathcal{F}_0$, as seen in \autoref{fig:F Evolution}. 

We note that the variation of the mass scale, $M$, also has a similar effect to varying $\mathcal{F}_0$, as the behaviour of $\mathcal{F}$ will depend on the relative size of $\mathcal{F}_0$ and $6H_0^2 \Omega_{\mathrm{de},0}/M^2$ from \eqref{Analytical F}. Similar to $\alpha$, varying $M$ will also change the domain of $\mathcal{F}$. It may seem that $M$ should not influence the evolution of $\mathcal{F}$ as it does not appear explicitly in \eqref{Background DE}. However, similar to $\alpha$, the $M$ dependence is hidden via $\mathcal{K} \propto M^{-2}$. Under a rescaling $\mathcal{K}\rightarrow M^2 \mathcal{K}$, we see that there is in fact a $M$ dependence in \eqref{Background DE}. However, if we instead work with the combination $M^2 \mathcal{F}$, then under the rescaling we find that \eqref{Background DE} becomes independent of $M$. Indeed, note that $\mathcal{F}$ appears as $M^2 \mathcal{F}$ in the Lagrangian \eqref{General Lagrangian} and from \eqref{Analytical F} we can write this as \begin{equation}
M^2 \mathcal{F} =  \left( M^2\mathcal{F}_0 + 6H_0^2\Omega_{\rm{de},0}\right) \left( \frac{\mathcal{K}}{\mathcal{K}_0}\right)^{1/2}-6H_0^2\Omega_{\rm{de},0}.
\end{equation} Hence, we see that any change in $M$ can be offset with a change in $\mathcal{F}_0$, i.e. $M$ and $\mathcal{F}_0$ are degenerate, as seen in \autoref{fig:F Evolution}. As expected, the choice of $M$ does not affect the functional form of $M^2\mathcal{F}$. We will therefore fix $M = H_0$, corresponding to the approximate mass scale dark energy begins to dominate, and keep $\mathcal{F}_0$ as a free parameter.

For solutions close to $w_{\mathrm{de}}=-1$, we do not expect to see large deviations from the analytical solution. Indeed, the previous discussion about the power law behaviour still holds for solutions with $w_{\mathrm{de}}$ sufficiently close to $-1$, as seen in \autoref{fig:F Evolution}. Although unfavoured by current observations, dark energy models with $w_{\mathrm{de}}\not=-1$ have not been completely ruled out and so we will allow for this in the subsequent analysis.

To summarise, for the background evolution we have 3 free parameters $\left\lbrace w_{\mathrm{de}},\Omega_{\mathrm{de},0},\mathcal{F}_0 \right\rbrace $ to specify, not 5, since $\alpha$ has no effect on the background evolution, other than a rescaling of the domain as a function of $\mathcal{K}$, and $M^2$ is degenerate with $\mathcal{F}_0$. While the background evolution only requires us to specify $\left\lbrace w_{\mathrm{de}},\Omega_{\mathrm{de},0},\mathcal{F}_0, M \right\rbrace $, as we will see in \autoref{sect:EoS}, at the level of linear perturbations the value of $\alpha$ and the other $\left\lbrace c_i\right\rbrace $ coefficients will be important.

\subsection{Sub-classes to Generalized Einstein-Aether}
There are a number of interesting sub-classes of the Generalized Einstein-Aether model that have been studied previously which we will mention here. 
\subsubsection{Linear Einstein-Aether}
The first is perhaps the most obvious simplification to this model, other than the absence of the Aether field, and that is to set $\mathcal{F}(\mathcal{K}) = \mathcal{K}$, and indeed this is the form of Einstein-Aether that was originally proposed in \cite{Mattingly1}.

In this case, the equations of motion become \begin{equation}
\nabla_\tau (J^{\tau}\hspace{0.1mm}_{\mu}) - c_4 A^\alpha \nabla_\alpha A^\nu \nabla_\mu A_\nu = \lambda A_\mu
\end{equation} and \begin{align}
U_{\alpha \beta} =& \hspace{1mm} \nabla_\mu \left( J_{(\alpha}\hspace{0.1mm}^{\mu} A_{\beta )} - J^\mu \hspace{0.1mm} _{(\alpha}A_{\beta)} - J_{(\alpha \beta)}A^\mu \right) + c_1 \left(\nabla_\mu A_\alpha \nabla^\mu A_\beta - \nabla_\alpha A_\mu \nabla_\beta A^\mu\right) \nonumber
\\& +c_4 A^\mu A^\nu \nabla_\mu A_\alpha \nabla_\nu A_\beta +\left(c_4 A^\mu A^\nu \nabla_\mu A^\tau \nabla_\nu A_\tau - A^\nu \nabla_\mu J^{ \mu}\hspace{0.1mm}_\nu  \right) A_\alpha A_\beta +\frac{1}{2} \mathcal{K} g_{\alpha \beta}.
\end{align} The energy density and pressure are then \begin{equation}\label{linear rho and P}
\rho_A = \frac{3}{2}\alpha H^2, \quad
P_A = - \frac{3}{2}\alpha H^2 - \alpha \dot{H}.
\end{equation} 

For a universe dominated by a fluid species with equation of state $P = w_i \rho$ the scale factor is 
$a \propto t^{2/3(1+w_i)}$. We therefore have that \begin{equation}
\frac{P_A}{\rho_A} = w_{\mathrm{de}} = -1 -\frac{2\dot{H}}{3H^2}=w_i
\end{equation} i.e. the equation of state parameter for Aether field in linear Einstein-Aether matches that of other fluids present in the universe \cite{Carroll}. This behaviour prevents linear Einstein-Aether, $\mathcal{F(\mathcal{K})=\mathcal{K}}$, from being a dark energy candidate and is one of the motivations for its generalization.

\subsubsection{Generalized Einstein-Aether with $c_4 = 0$}
As already mentioned, many previous studies of Einstein-Aether models set $c_4 = 0$. It is often argued that this can be done via a redefinition of the coefficients. However, we will see in the next section that this can only be achieved after a specific choice of $A^\mu$ which has further consequences at the level of linear perturbations. In this case, the equations of motion become  \begin{equation} 
\nabla_\nu (\mathcal{F}_\mathcal{K}J^{\nu}\hspace{0.1mm}_{\mu})  = \lambda A_\mu,
\end{equation} and \begin{align} 
U_{\alpha \beta} =& \hspace{1mm}\nabla_\mu \left( \mathcal{F}_{\mathcal{K}} \left[ J_{(\alpha}\hspace{0.1mm}^{\mu} A_{\beta )} - J^\mu \hspace{0.1mm} _{(\alpha}A_{\beta)} - J_{(\alpha \beta)}A^\mu \right) \right] + c_1 \mathcal{F}_{\mathcal{K}} \left(\nabla_\mu A_\alpha \nabla^\mu A_\beta - \nabla_\alpha A_\mu \nabla_\beta A^\mu\right) \nonumber \\& - A_\alpha A_\beta A^\nu \nabla_\mu (\mathcal{F}_{\mathcal{K}} J^{ \mu}\hspace{0.1mm}_\nu ) +\frac{1}{2} M^2 \mathcal{F} g_{\alpha \beta}.
\end{align}

\subsubsection{The Khronometric model}
The Khronometric model \cite{Blas1,Blas2} is a version of Einstein-Aether where the Aether field is constrained via a scalar field, $\varphi$, called the Khronon. In this case, the field is defined as \begin{equation} \label{Khronon}
A_\mu = -\frac{\partial_\mu \varphi}{\sqrt{-g^{\alpha \beta} \partial_\alpha \varphi \partial_\beta \varphi}},
\end{equation} and so the time-like unit norm constraint is satisfied automatically. In doing so, the Aether is restricted to be orthogonal to a set of space-like surfaces defined by $\varphi$. At background order we assume $\varphi = \varphi(t)$ and so from \eqref{Khronon} we have that 
$A^\mu = (1,0,0,0)$, which is the same as before. Therefore, the choice of the Khronon definition has no effect on background dynamics.

The khronometric model was first proposed in \cite{Blas1}, where $\varphi$ sets a preferred global time coordinate. It was discussed how this model describes the low energy limit of the consistent extension of Horava gravity, a quantum theory of gravity. At low energies, this reduces to a Lorentz-violating scalar-tensor gravity theory. For more details see \cite{Blas1,Blas2,Horava1,Horava2}.

For this choice of the Aether field, the $c_1, c_3$ and $c_4$ terms are no longer independent. The twist vector is defined as \cite{Eling} \begin{equation}
\omega _\alpha = \varepsilon _{\alpha \beta \mu \nu} A^\beta \nabla^\mu A^\nu,
\end{equation} where $\varepsilon _{\alpha \beta \mu \nu}$ is the 4-dimensional Levi-Civita symbol, and $\omega_\alpha = 0$ if $A^\mu$ is hypersurface orthogonal. If $\omega_\alpha = 0$ then\begin{equation}
w^\alpha w_\alpha = 0 = \varepsilon _{\alpha \beta \mu \nu}  \varepsilon ^{\alpha \gamma \rho \sigma }A^\beta A_\gamma \nabla^\mu A^\nu \nabla_\rho A_\sigma = - \delta ^{\gamma \rho \sigma} _{\beta \mu \nu}A^\beta A_\gamma \nabla^\mu A^\nu \nabla_\rho A_\sigma,
\end{equation} where $\delta ^{\gamma \rho \sigma} _{\beta \mu \nu}$ is the generalized Kronecker delta. Therefore, \begin{align}
&-A^\gamma A_\gamma \nabla_\rho A_\sigma \nabla^\rho A^\sigma - A^\sigma A_\gamma \nabla _\rho A_\sigma \nabla^\gamma A^\rho - A^\rho A_\gamma \nabla_\rho A_\sigma \nabla^\sigma A^\gamma \nonumber \\
&+A^\gamma A_\gamma \nabla_\rho A_\sigma \nabla^\sigma A^\rho + A^\rho A_\gamma \nabla_\rho A_\sigma \nabla^\gamma A^\sigma + A^\sigma A_\gamma \nabla_\rho A_\sigma \nabla^\rho A^\gamma = 0.
\end{align} From $A_\gamma \nabla^\rho A^\gamma = \nabla^\rho (A_\gamma A^\gamma) - A^\gamma \nabla^\rho A_\gamma$, applying the unit norm constraint gives $A_\gamma \nabla^\rho A^\gamma =  0$, and so \begin{equation} \label{coefficient redef}
 A^\rho A_\gamma \nabla_\rho A_\sigma \nabla^\gamma A^\sigma = \nabla_\rho A_\sigma \nabla^\sigma A^\rho -\nabla_\rho A_\sigma \nabla^\rho A^\sigma.
\end{equation} Note that the left-hand side of \eqref{coefficient redef} is the $c_4$ term in \eqref{Kinetic Tensor}. Since the terms on the right-hand side of \eqref{coefficient redef} are related to the $c_1$ and $c_3$ terms, we are able to absorb $c_4$ into the other coefficients effectively setting $c_4=0$ i.e. $c_1 \rightarrow c_1'=c_1-c_4$ and $c_3 \rightarrow c_3'=c_3+c_4$ giving \begin{equation} 
K^{\alpha \beta} \hspace{0.1mm} _{\mu \nu} = c_1' g^{\alpha \beta} g_{\mu \nu} + c_2\delta^\alpha _\mu \delta^\beta _\nu + c_3'\delta^\alpha _\nu \delta^\beta _\mu.
\end{equation} We therefore see that it is possible to set $c_4=0$, but only if the choice is made that $A^\mu$ is also hypersurface orthogonal. While this has no effect at background order, we will see later that differences arise at the level of linear perturbations for the vector sector. Furthermore, this is not the only choice we can make as \eqref{coefficient redef} also allows a redefinition which could remove $c_1$ or $c_3$ instead.

\section{Linear perturbations} \label{sect:Pert}
We will present results for perturbations in the scalar sector in both the synchronous and conformal Newtonian gauge. We perturb the metric as \begin{equation}
g_{\mu \nu} =  \bar{g}_{\mu \nu} + \delta g_{\mu \nu} = a^2(\tau) (\eta_{\mu \nu} + h_{\mu \nu}),
\end{equation} such that \begin{equation}
ds^2 = a^2(\tau)\left[ -( 1 + 2\Psi) d\tau^2 + (\delta_{ij} + h_{ij})dx^i dx^j \right],
\end{equation} where we now work in conformal time, $\tau$. In the synchronous gauge we set $\Psi = 0$ and decompose $h_{ij}$ into \cite{CosPert,DEAni} \begin{equation} \label{Synchronous decomp}
h_{ij} = \hat{k}_i \hat{k}_j h + \left(  \hat{k}_i \hat{k}_j - \frac{1}{3}\delta_{ij} \right) 6 \eta + 2\hat{k}_{(i} \left( h ^{V1} \hat{l}_{j)} + h^{V2}\hat{m}_{j)}  \right) \nonumber\\
+ h ^{+}\left( \hat{l}_i \hat{l}_j - \hat{m}_i \hat{m}_j   \right) +h ^{\times}\left( \hat{l}_i \hat{m}_j -  \hat{l}_j  \hat{m}_i \right),
\end{equation} where the unit vectors $\left\lbrace \hat{k}, \hat{l}, \hat{m} \right\rbrace $ form an orthonormal basis in $k$-space. Here, $h$ and $\eta$ are the scalar perturbations, $h^{V1}$ and $h^{V2}$ are the vector perturbations, and $h^{+}$ and $h^\times$ are the tensor perturbations. In the conformal Newtonian gauge we set $h^{\mathrm{scalar}}_{ij} = -2\Phi \delta_{ij}$, while the vector and tensor perturbations are as before in the synchronous gauge. 

We also perturb the Aether field as \cite{Ferreira3}\begin{equation}
A^\mu = \bar{A}^\mu + \delta A^\mu =  \frac{1}{a} (1 + X, \partial ^i V+\mathrm{i}B^i),
\end{equation} where $V$ is the longitudinal scalar mode and $B^i$ is the transverse vector mode such that $k_i B^i = 0$. The unit norm constraint demands that  $X = -\Psi$ and so \begin{equation} \label{Perturbed Aether Vector}
\delta A^\mu = \frac{1}{a}(-\Psi, \partial^i V+\mathrm{i}B^i).
\end{equation} Hence, we see that the time-like unit norm constraint means that there is only one scalar degree of freedom, $V$, along with a transverse vector mode, $B^i$. In what follows, we will suppress over-bars to denote background order quantities.

The perturbed energy momentum tensor is given by \begin{align} \label{Perturbed Aether EMT}
\delta U_{\alpha \beta} = &\,\delta\left( \nabla_\mu\left[ \mathcal{F}_{\mathcal{K}} \left( J_{(\alpha}\hspace{0.1mm}^{\mu} A_{\beta )} - J^\mu \hspace{0.1mm} _{(\alpha}A_{\beta)} - J_{(\alpha \beta)}A^\mu \right) \right]  \right) \nonumber  \\
&+ c_1 \mathcal{F}_{\mathcal{KK}} \delta \mathcal{K}\left( \nabla_\mu A_\alpha \nabla^\mu A_\beta - \nabla_\alpha A_\mu \nabla_\beta A^\mu \right) +c_1 \mathcal{F}_{\mathcal{K}} \delta \left( \nabla_\mu A_\alpha \nabla^\mu A_\beta - \nabla_\alpha A_\mu \nabla_\beta A^\mu \right) \nonumber \\
& +c_4 \mathcal{F}_{\mathcal{KK}} \delta \mathcal{K} A^\mu A^\nu \nabla_\mu A_\alpha \nabla_\nu A_\beta +c_4 \mathcal{F}_{\mathcal{K}} \delta \left( A^\mu A^\nu \nabla_\mu A_\alpha \nabla_\nu A_\beta\right)  \nonumber
\\& + \delta\left(  \left[c_4 \mathcal{F}_{\mathcal{K}}A^\mu A^\nu \nabla_\mu A^\tau \nabla_\nu A_\tau - A^\nu \nabla_\mu (\mathcal{F}_{\mathcal{K}} J^{ \mu}\hspace{0.1mm}_\nu ) \right] A_\alpha A_\beta\right) \nonumber \\
&+\frac{1}{2} M^2 \left( \mathcal{F}\delta g_{\alpha \beta} +g_{\alpha \beta} \mathcal{F}_{\mathcal{K}}  \delta \mathcal{K}\right) . 
\end{align} For a general energy-momentum tensor, $E_{\mu \nu}$, we can decompose its perturbations as \cite{CosPert}\begin{equation}
\delta E^\mu\hspace{0.1mm}_{\nu} = (\delta \rho + \delta P) u^\mu u_\nu + \delta P \delta^\mu \hspace{0.1mm}_{ \nu} +(\rho + P)(\delta u^\mu u_\nu + \delta u_\nu u^\mu) + P \Pi^\mu\hspace{0.1mm}_{\nu},
\end{equation} where $u^\mu = \frac{1}{a}(1,0,0,0)$, $\delta u^\mu = \frac{1}{a}(0, v^i)$ and $\Pi^\mu\hspace{0.1mm} _\nu$ is the anisotropic stress, with the properties $u^\nu \Pi^\mu\hspace{0.1mm}_\nu=0$, $\Pi^\mu\hspace{0.1mm}_\nu= \Pi_\nu\hspace{0.1mm}^\mu$, and $\Pi^\mu\hspace{0.1mm}_\mu=0$. Projecting out the perturbed fluid variables, we find that \begin{align}
\delta E^0\hspace{0.1mm}_{0} &= - \delta \rho, \\
\delta E^0\hspace{0.1mm}_{i} &= (\rho + P)v_i, \\
\delta E^i\hspace{0.1mm}_{0} &= -(\rho + P)v_i, \\ \label{ij projection}
\delta E^i\hspace{0.1mm}_{j} &= P\Pi^i _{j} + \delta P \delta^i \hspace{0.1mm}_{j}.
\end{align} Similar to $h_{ij}$, we can decompose $v_i$ and $\Pi _{ij}$ into scalar, vector, and tensor parts. They are given by \cite{Moss} \begin{align}
&v_i = V^S \hat{k}_i + V^{V1} \hat{l}_i + V^{V2} \hat{m}_i, \\
\Pi_{ij} = \left( \hat{k}_i \hat{k}_j - \frac{1}{3}\delta_{ij} \right) \Pi ^S + 2\hat{k}_{(i} &\left( \Pi ^{V1} \hat{l}_{j)} + \Pi^{V2}\hat{m}_{j)}  \right) + \Pi ^{+}\left( \hat{l}_i \hat{l}_j - \hat{m}_i \hat{m}_j   \right) +\Pi ^{\times}\left( \hat{l}_i \hat{m}_j -  \hat{l}_j  \hat{m}_i \right),
\end{align} whereas the transverse vector, $B_i$, only has vector modes i.e. \begin{equation}
B_i = B^{V1} \hat{l}_i + B^{V2}\hat{m}_i.
\end{equation}

In a general gauge, the perturbed fluid variables from \eqref{Perturbed Aether EMT} in $k$-space are then \begin{align} \label{delta rho}
a^2\delta \rho &=  \alpha\left[ 3\mathcal{F}_\mathcal{KK} \delta \mathcal{K} \mathcal{H}^2 + \mathcal{F}_\mathcal{K} \mathcal{H}\left( \frac{1}{2}h' - k^2V - 3\mathcal{H}\Psi \right)\right]  + c_{14} \mathcal{F}_\mathcal{K} k^2 (V' + \mathcal{H}V + \Psi), \\ \nonumber \\ \label{delta P}
a^2 \delta P &=  \alpha \mathcal{F}_{\mathcal{K}} \left[ \mathcal{H}\Psi' + \left( 2 \mathcal{H}' + \mathcal{H}^2\right) \Psi -\frac{1}{6} \left( h'' + 2\mathcal{H}h' \right) +\frac{1}{3}k^2 \left( V' + 2\mathcal{H}V \right)\right]  \nonumber \\
&-\alpha \mathcal{F}_{\mathcal{KK}} \left[ \left( \mathcal{H}' +  2 \mathcal{H}^2+ \frac{\mathcal{F}_{\mathcal{KKK}}}{\mathcal{F}_{\mathcal{KK}}}\mathcal{K}' \mathcal{H}\right) \delta K + \delta \mathcal{K}' \mathcal{H}  -\frac{1}{6}\mathcal{K}' \left(  12\mathcal{H} \Psi  + 2k^2V - h' \right) \right], \\
\nonumber \\
a^2(\rho + P) v_i &=  \mathrm{i}\alpha \left[ \left( \mathcal{F}_{\mathcal{K}}  \left( \mathcal{H}^2  - \mathcal{H}' \right) - \mathcal{F}_{\mathcal{KK}}\mathcal{K}' \mathcal{H} \right)\xi_i-\frac{1}{2}k^2B_i \right] +\mathrm{i}\left( \frac{3}{2}c_2+c_1\right) \mathcal{F}_\mathcal{K}k^2B_i\nonumber \\
&+\mathrm{i} c_{14}\left[  \mathcal{F}_{\mathcal{K}} \left( \xi''_i + 2\mathcal{H}\xi'_i + \left( \mathcal{H}'  + \mathcal{H}^2\right) \xi_i +k_i\Psi' +\mathcal{H} k_i\Psi \right) + \mathcal{F}_{\mathcal{KK}}\mathcal{K}' \left( \xi'_i+\mathcal{H}\xi_i +k_i\Psi \right)\right], \\ \nonumber \\
a^2 P\Pi ^i \hspace{0.1mm}_j &= c_{13} \left[ \mathcal{F}_{\mathcal{KK}} \mathcal{K}' \left( k^i k_j V - \frac{1}{2}h^i \hspace{0.1mm}_j \hspace{0.1mm}'\right)  + \mathcal{F}_{\mathcal{K}} k^i k_j (V' +2\mathcal{H}V) - \mathcal{F}_\mathcal{K}\left( \frac{1}{2}h^i \hspace{0.1mm}_j \hspace{0.1mm}''+\mathcal{H} h^i \hspace{0.1mm} _j \hspace{0.1mm}'\right)  \right. \nonumber \\
&+\frac{1}{6}\left( \mathcal{F}_{\mathcal{KK}} \mathcal{K}' \left(h' - 2k^2 V \right) +\mathcal{F}_\mathcal{K} \left( h'' + 2\mathcal{H}h' \right)   - 2\mathcal{F}_\mathcal{K}k^2 (V' +2 \mathcal{H}V) \right) \delta ^i\hspace{0.1mm}_j \nonumber \\
&+\left. \left(\mathcal{F}_{\mathcal{K}} \mathcal{H} +\frac{1}{2}\mathcal{F}_{\mathcal{KK}}\mathcal{K}'\right) \left( k^i B_j + k_j B^i   \right) +\frac{1}{2} \mathcal{F}_{\mathcal{K}}\left( k^i B_j\hspace{0.1mm}' +k_j  B^i\hspace{0.1mm}' \right)\right], 
\end{align} where primes denote conformal time differentiation, $c_{13} = c_1 +c_3$, $c_{14}=c_1-c_4$, $c_{123}= c_1+c_3+c_3$, and $\xi_i=k_iV+B_i$.

\subsection{Scalar sector} \label{sect:ScalarPert}
The scalar components of $v_i$ and $\Pi^i\hspace{0.1mm}_j$ are obtained via $V^S=\hat{k}^i v_i$ and $\Pi ^S = \frac{3}{2}\left( \hat{k}_i \hat{k}^j - \frac{1}{3}\delta^{j} _i \right) \Pi^i _{j}$. If we further define $\theta ^S = \mathrm{i}V^S/k=\mathrm{i}k^i v_i/k^2$, then we have that \begin{align} a^2(\rho+ P) \theta ^S &=  \alpha \left[\mathcal{F}_{\mathcal{K}}\left(\mathcal{H}'-\mathcal{H}^2 \right) + \mathcal{F}_{\mathcal{KK}}\mathcal{K}'\mathcal{H}\right]V\nonumber \\
	&- c_{14}\left[  \mathcal{F}_{\mathcal{K}} \left( V'' + 2\mathcal{H}V' + \left( \mathcal{H}'  + \mathcal{H}^2\right) V +\Psi' +\mathcal{H} \Psi \right) + \mathcal{F}_{\mathcal{KK}}\mathcal{K}' \left( V'+\mathcal{H}V+\Psi \right)\right],\\ \nonumber \\ \frac{2}{3}a^2P\Pi^S = c_{13} &\left(\hat{k}_i \hat{k}^j-\frac{1}{3}\delta^{j} _i \right)\left[\mathcal{F}_{\mathcal{KK}} \mathcal{K}'\left(k^i k_jV-\frac{1}{2}h'^i\hspace{0.1mm}_j\right)+\mathcal{F}_{\mathcal{K}}k^ik_j(V'+2\mathcal{H}V)-\mathcal{F}_\mathcal{K}\left(\frac{1}{2}h''^i\hspace{0.1mm}_j+\mathcal{H}h'^i\hspace{0.1mm}_j\right)\right].
\end{align} Note that the expression for $\Pi^S$ will simplify further once we specify the gauge. We further define the entropy perturbation, $\Gamma$, as \begin{equation}
	w \Gamma = \left(\frac{\delta P}{\delta \rho} - \frac{dP}{d \rho} \right)\delta.
\end{equation} It should be noted that whatever gauge we choose to work in, both $\Pi^S$ and $\Gamma$ are gauge invariant.
The perturbed Aether field equation of motion is obtained from perturbing \eqref{Vector EoM}. Taking the $i$-component, the $\hat{k}^i$ direction will yield the equation of motion governing the perturbation $V$, given by \begin{align} \label{Perturbed V} &c_1 \left[ V'' +2\mathcal{H}V' + (2 \mathcal{H}^2 +k^2) V +\Psi' +2\mathcal{H}\Psi - \frac{1}{2} \hat{k}^i \hat{k}_j h^j \hspace{0.1mm}_i\hspace{0.1mm}' \right] +c_2\left[ \left( k^2 + 3\mathcal{H}^2 -3\mathcal{H}'\right) V +3\mathcal{H}\Psi - \frac{1}{2}h' \right]  \nonumber \\
+ \hspace{1mm}&c_3  \left[ \left( k^2 +\mathcal{H}^2 - \mathcal{H}' \right) V +\mathcal{H}\Psi  - \frac{1}{2} \hat{k}^i \hat{k}_j h^j \hspace{0.1mm}_i\hspace{0.1mm}' \right] -c_4 \left[V'' +2\mathcal{H}V' + (\mathcal{H}' +\mathcal{H}^2) V +\Psi' +\mathcal{H}\Psi  \right] 
\nonumber \\
-&\frac{\mathcal{F}_\mathcal{KK}}{\mathcal{F}_\mathcal{K}}\left( \alpha \delta\mathcal{K\mathcal{H}} +\mathcal{K}' \left[ \alpha \mathcal{H}V  -c_{14}(V' + \mathcal{H}V + \Psi) \right] \right) = 0,
\end{align} where we have substituted in for $\lambda$. 

\subsubsection{Conformal Newtonian gauge}
In the conformal Newtonian gauge, where the metric perturbations are parametrized via $\Psi$ and $\Phi$, we have that  \begin{align}
a^2\delta \rho &=  \left[3 \mathcal{F}_\mathcal{KK} \delta \mathcal{K} \mathcal{H}^2 - \mathcal{F}_\mathcal{K} \mathcal{H}\left(  k^2V + 3\mathcal{H}\Psi + 3\Phi ' \right) \right] + c_{14} \mathcal{F}_\mathcal{K} k^2 (V' + \mathcal{H}V + \Psi), \\ \nonumber \\
a^2 \delta P &=  \alpha \mathcal{F}_{\mathcal{K}} \left[ \mathcal{H}\Psi' + \left( 2 \mathcal{H}' + \mathcal{H}^2\right) \Psi + \Phi'' +2\mathcal{H}\Phi' +\frac{1}{3}k^2 \left( V' + 2\mathcal{H}V \right)\right] \nonumber \\ &-\alpha \mathcal{F}_{\mathcal{KK}} \left[ \left( \mathcal{H}' +  2 \mathcal{H}^2+ \frac{\mathcal{F}_{\mathcal{KKK}}}{\mathcal{F}_{\mathcal{KK}}}\mathcal{K}' \mathcal{H}\right) \delta K + \delta \mathcal{K}' \mathcal{H}  -\frac{1}{3}\mathcal{K}' \left(  6\mathcal{H} \Psi+3\Phi'  +k^2V \right) \right], \\\nonumber\\\label{theta CN}
a^2 (\rho+ P) \theta ^S&=  \alpha \left[ \mathcal{F}_{\mathcal{K}}  \left( \mathcal{H}'  - \mathcal{H}^2 \right) + \mathcal{F}_{\mathcal{KK}}\mathcal{K}' \mathcal{H} \right]V \nonumber \\
&- c_{14}\left[  \mathcal{F}_{\mathcal{K}} \left( V'' + 2\mathcal{H}V' + \left( \mathcal{H}'  + \mathcal{H}^2\right) V +\Psi' +\mathcal{H} \Psi \right) + \mathcal{F}_{\mathcal{KK}}\mathcal{K}' \left( V'+\mathcal{H}V+\Psi \right)\right], \\ \nonumber \\ \label{CN Pi}
a^2P\Pi^S &= c_{13} \left[ \mathcal{F}_{\mathcal{KK}} \mathcal{K}' k^2 V  + \mathcal{F}_{\mathcal{K}} k^2(V' +2\mathcal{H}V) \right]. 
\end{align} The perturbed Aether field equation of motion reads \begin{align}  \label{Vector EoM CN}
\alpha &\left[ \left( \mathcal{H}^2 - \mathcal{H}' +k^2\right)V +\mathcal{H}\Psi +\Phi' - \frac{\mathcal{F}_\mathcal{KK}}{\mathcal{F}_\mathcal{K}}\left( \delta \mathcal{K}\mathcal{H}+\mathcal{K}'\mathcal{H}V \right) \right] \nonumber \\
+\,c_{14} &\left[ V'' +2\mathcal{H}V' + (\mathcal{H}^2 +\mathcal{H}') V +\Psi' +\mathcal{H}\Psi + \frac{\mathcal{F}_\mathcal{KK}}{\mathcal{F}_\mathcal{K}}\mathcal{K}' (V' + \mathcal{H}V + \Psi) \right] - 2c_2 k^2 V =0.
\end{align}

\subsubsection{Synchronous gauge}
In the synchronous gauge, where $h_{ij}$ is decomposed into $h$ and $\eta$ as in \eqref{Synchronous decomp}, we find that \begin{align}
a^2\delta \rho &=  \alpha\left[ 3\mathcal{F}_\mathcal{KK} \delta \mathcal{K} \mathcal{H}^2 + \mathcal{F}_\mathcal{K} \mathcal{H}\left( \frac{1}{2}h' - k^2V \right)\right]  +c_{14} \mathcal{F}_\mathcal{K} k^2 (V' + \mathcal{H}V) \\ \nonumber \\
a^2 \delta P &= \frac{1}{3}\alpha \mathcal{F}_{\mathcal{K}} \left[k^2 \left( V' + 2\mathcal{H}V \right) - \frac{1}{2}h'' - \mathcal{H}h' \right] \nonumber \\ &-\alpha \mathcal{F}_{\mathcal{KK}} \left[ \left( \mathcal{H}' +  2 \mathcal{H}^2+ \frac{\mathcal{F}_{\mathcal{KKK}}}{\mathcal{F}_{\mathcal{KK}}}\mathcal{K}' \mathcal{H}\right) \delta K + \delta \mathcal{K}' \mathcal{H}  -\frac{1}{6}\mathcal{K}' \left(h'  +2k^2V \right)\right],
\\\nonumber \\
a^2 (\rho+ P) \theta ^S&=  \alpha \left[ \mathcal{F}_{\mathcal{K}}  \left( \mathcal{H}'  - \mathcal{H}^2 \right) + \mathcal{F}_{\mathcal{KK}}\mathcal{K}' \mathcal{H} \right]V \nonumber \\
&- c_{14}\left[  \mathcal{F}_{\mathcal{K}} \left( V'' + 2\mathcal{H}V' + \left( \mathcal{H}'  + \mathcal{H}^2\right) V \right) + \mathcal{F}_{\mathcal{KK}}\mathcal{K}' \left( V'+\mathcal{H}V \right)\right], \\ \nonumber \\ \label{Sync Pi}
a^2P\Pi^S &= c_{13}  \left[ \mathcal{F}_{\mathcal{KK}} \mathcal{K}' \left( k^2V - \frac{1}{2}\left( h +6 \eta \right) \right)  + \mathcal{F}_{\mathcal{K}} k^2(V' +2\mathcal{H}V) \right. \nonumber \\
&\left. - \mathcal{F}_\mathcal{K}\left( \frac{1}{2}\left( h'' +6 \eta'' \right)+\mathcal{H} \left( h' +6 \eta' \right)\right)  \right].
\end{align} The perturbed equation of motion for the Aether field reads \begin{align} 
\alpha &\left[(\mathcal{H}^2 -\mathcal{H}' +k^2 )V -\frac{1}{2}\left( h' + 4\eta ' \right) -\frac{\mathcal{F}_\mathcal{KK}}{\mathcal{F}_\mathcal{K}} \left( \delta\mathcal{K}\mathcal{H} +\mathcal{K}'\mathcal{H}V \right) \right] \nonumber\\
+\,c_{14} &\left[ V'' +2\mathcal{H}V' + (\mathcal{H}^2 +\mathcal{H}') +\frac{\mathcal{F}_\mathcal{KK}}{\mathcal{F}_\mathcal{K}} \mathcal{K}'(V'+\mathcal{H}V)\right] + c_2 \left(h' +6\eta' - 2k^2V \right)=0.
\end{align}

\subsection{Vector and tensor sectors}
In the vector and tensor sectors, the vector and tensor modes of $v_i$ and $\Pi^i\hspace{0.1mm}_j$ can be computed via $V^{V1}=\hat{l}^iv_i$, $\Pi^{V1} =\hat{k}_i\hat{l}^j\Pi^i _j$, and $ \displaystyle \Pi ^+=\frac{1}{2}\left( \hat{l}_i\hat{l}^j -  \hat{m}_i\hat{m}^j  \right) \Pi ^i _{j} $. Equivalent expressions also exist for the $V2$ modes and $\Pi^\times$. Also, analogous to $\theta^S$, we can define $\theta^{V1} = \mathrm{i}V^{V1}/k =\mathrm{i}\hat{l}^iv_i/k$ and so we have that \begin{align} \label{Theta V}
a^2(\rho + P)k\theta^{V1}  &= \alpha\left[ \mathcal{F}_{\mathcal{K}}\left( \mathcal{H}' - \mathcal{H}^2 \right)+\mathcal{F}_{\mathcal{KK}} \mathcal{K}' \mathcal{H} \right]  B^{V1} +\frac{1}{2}(c_3-c_1) \mathcal{F}_\mathcal{K}k^2 B^{V1} \nonumber \\
&-c_{14} \left[\mathcal{F}_{\mathcal{K}}\left(  B^{V1''} + 2\mathcal{H} B^{V1 '} + \left( \mathcal{H}' +\mathcal{H}^2 \right) B^{V1}\right)  + \mathcal{F}_{\mathcal{KK}} \mathcal{K}'  \left( B^{V1'} +\mathcal{H} B^{V1} \right) \right], \\ \nonumber \\
\label{PiV}
a^2P\Pi^{V1} &= c_{13}\left[ \frac{1}{2} \mathcal{F}_{\mathcal{K}}\left( k B^{V1'} - h^{V1''}  \right)+ \left(\mathcal{F}_{\mathcal{K}}\mathcal{H} +\frac{1}{2}\mathcal{F}_{\mathcal{KK}}\mathcal{K}'\right) \left( k B^{V1} -  h^{V1'}  \right) \right] , \\ \nonumber \\ \label{h plus}
 a^2P\Pi ^+ &= -c_{13} \left[\frac{1}{2} \mathcal{F}_{\mathcal{K}} h^{+''}+ \left(\mathcal{F}_{\mathcal{K}} \mathcal{H}+\frac{1}{2}\mathcal{F}_{\mathcal{KK}}\mathcal{K}'\right)h^{+'} \right] .
\end{align} The time-time and traced $ij$-components are zero in the vector and tensor sectors since $\delta \rho$ and $\delta P$ only have scalar modes.

The equation of motion for the Aether field in the $\hat{l}^i$ direction is given by \begin{align} \label{PertVectorEoMVS}
\alpha&\left[ \left( \mathcal{H}^2 - \mathcal{H}' \right) B^{V1} - \frac{1}{2}k h^{V1'}  - \frac{\mathcal{F}_{\mathcal{KK}}}{\mathcal{F}_{\mathcal{K}}}\mathcal{K}'\mathcal{H}B^{V1}\right] + c_1 k^2 B^{V1} +\frac{3}{2}c_2 kh^{V1'} \nonumber \\
+\,c_{14} &\left[  B^{V1''} + 2 \mathcal{H} B ^{V1'} + \left(\mathcal{H}' + \mathcal{H}^2 \right) B^{V1} + \frac{\mathcal{F}_{\mathcal{KK}}}{\mathcal{F}_{\mathcal{K}}}\mathcal{K}'  \left( B^{V1'} + \mathcal{H} B^{V1} \right) \right] =0.
\end{align} Note that the two vector and tensor modes are interchangeable. From here on we will not discriminate between them and denote them simply as $\theta^V$, $\Pi^V$ and $\Pi^T$.

\subsection{Vector modes in the Khronon}

If we restrict ourselves to the case where the Aether field is defined by the Khronon in \eqref{Khronon}, then we find that
\begin{equation}
\delta A_\mu = \frac{a}{\varphi'} \left[ -\partial_\mu \delta \varphi  + \partial_\mu \varphi \left( \Psi + \frac{\delta \varphi'}{\varphi'} \right) \right],
\end{equation} where $\delta \varphi $ is the perturbed Khronon field. The time component is then $\delta A_0 = a\Psi$, which is a consequence of the time-like unit norm constraint, as in \eqref{Perturbed Aether Vector}. However, if we calculate the spatial component we find that  \begin{equation}
\delta A_i = -\frac{a}{\varphi'}\partial_i \delta \varphi  \Rightarrow B_i = 0
\end{equation} i.e. there is no propagating transverse vector mode. Therefore, if we redefine $\dfrac{1}{\varphi'}\partial_i \delta \varphi  = \partial_i V$ then we obtain the results from \autoref{sect:ScalarPert}. Therefore, the scalar sector for Generalized Einstein-Aether and the Khronon are completely equivalent \cite{Blas2}, up to a redefinition of the coefficients discussed previously.

%\color{red}{\textbf{Can this be shown by thinking about $c_i$? Possibly. Khronon definition allows either $c_1,c_3$ or $c_4$ to be effectively set to zero. (Most people set $c_4=0$). But only the case of setting $c_1=0$ means that the vector modes do not propagate, since the sound speed becomes zero. See \eqref{Wave B}.}}\color{black}

\section{Equations of state for perturbations} \label{sect:EoS}
\subsection{Scalar sector}

%The perturbed conservation equation, $\delta (\nabla_\mu T^\mu_\nu)=0$, gives two evolution equations for the density perturbation, $\delta \rho$, and velocity field, $\theta^S$. For example, in the synchronous gauge they are given by \begin{equation} \label{ConsEq1}
%\left(\frac{\delta}{1+w} \right)' = k^2 \theta^S - \frac{1}{2}h' - \frac{3\mathcal{H}w}{1+w}\Gamma 
%\end{equation} and \begin{equation} \label{ConsEq2}
%(1+w)\theta^{S'} = \mathcal{H}(1+w)\left( 3\frac{dP}{d\rho}-1 \right)\theta^S - \frac{dP}{d\rho}\delta - w\Gamma +\frac{2}{3}w \Pi^S. 
%\end{equation} The metric perturbations are evolved via Einstein's equation. However, the evolution of $\Pi^S$ and $\Gamma$ are not known. Hence, \eqref{ConsEq1} and \eqref{ConsEq2} are not closed. If we can somehow specify the gauge invariant entropy perturbation, $\Gamma$, and anisotropic stress, $\Pi^{S}$.  these equations close. This is the essence of the EoS approach. The equations of state for perturbations are obtained by fully eliminating the degrees of freedom, $V$, $V'$ and $V''$. In doing so, $\Gamma$ and $\Pi^S$ are written in terms of the perturbed fluid variables, metric perturbations, and derivatives only. Our approach is to eliminate the degrees of freedom via our expressions for $\delta \rho$ and $\theta^S$. Once these have been obtained, the degrees of freedom can be eliminated from $\Pi^S$, $\delta P$, and hence $\Gamma$.
We now derive the equations of state, $\Gamma$ and $\Pi^{S,V,T}$, in terms of the other perturbation variables by fully eliminating the internal degrees of freedom introduced by the theory i.e. $V$, $B^i$, and their derivatives. In the scalar sector we do this via the expressions for $\delta \rho$ and $\theta^S$. Let us first work in the conformal Newtonian gauge. Initially it may not seem possible to eliminate the degrees of freedom as we have that $\theta^S \equiv \theta^S (V,V',V'')$ and $\delta \rho \equiv \delta \rho(V,V')$, i.e. we have three unknowns and only two equations. However, we can use the perturbed Aether field equation of motion \eqref{Vector EoM CN} to reduce the dimensionality of the problem. Using this to eliminate $V''$ in \eqref{theta CN} and gathering terms in $V$ and $V'$, we find that \begin{align}
a^2 \delta \rho =& \hspace{1mm} c_{14}\mathcal{F}_\mathcal{K}k^2 V'  - \left[\alpha \mathcal{F}_\mathcal{K} -c_{14}\mathcal{F}_\mathcal{K} + \frac{6 \alpha^2 \mathcal{F}_\mathcal{KK} \mathcal{H}^2}{a^2 M^2} \right] \mathcal{H}k^2 V \nonumber \\
 +\hspace{1mm}&c_{14}\mathcal{F}_\mathcal{K} k^2 \Psi - 3\alpha \mathcal{H}\left(\mathcal{F}_\mathcal{K} + \frac{6\alpha \mathcal{F}_\mathcal{KK}\mathcal{H}^2}{a^2 M^2} \right) (\mathcal{H}\Psi + \Phi'), \\ \label{ThetaS_Rearranged}
a^2\rho(1+w_{\mathrm{de}}) \theta^S =& \left[ c_{123} \mathcal{F}_\mathcal{K} +\frac{2\alpha^2 \mathcal{H}^2 \mathcal{F}_\mathcal{KK}}{a^2 M^2} \right]k^2 V +\alpha\left( \mathcal{F}_\mathcal{K} + \frac{6\alpha \mathcal{H}^2 \mathcal{F}_\mathcal{KK}}{a^2 M^2} \right)(\mathcal{H}\Psi +\Phi'),
\end{align} where we have substituted in for $\mathcal{K}$ from \eqref{Scalar K} and \begin{equation}
\delta \mathcal{K} = -\frac{2 \alpha\mathcal{H}}{a^2 M^2}(k^2 V + 3\mathcal{H}\Psi +3\Phi').
\end{equation} So we see that in fact $\theta^S \equiv \theta^S(V)$. Note that we can already see the emergence of the gauge invariant combination, $\mathcal{H}\Psi + \Phi'$, in the $0i$-component of Einstein's equation that was used in \cite{PPF1,PPF2,PPF3}.

We can then write this system of equations as \begin{equation} \label{activation aether}
a^2\begin{pmatrix}
\delta\rho\\
\rho (1+w_{\mathrm{de}})\theta^S
\end{pmatrix} = k^2\begin{pmatrix}
A & B \\
0 & C
\end{pmatrix} \begin{pmatrix}
V'\\
V
\end{pmatrix} + 	\begin{pmatrix}
D\\
E
\end{pmatrix},
\end{equation}  with \begin{align}
A &= c_{14}\mathcal{F}_\mathcal{K}, \\
B &= \left[c_{14}\mathcal{F}_\mathcal{K}-\alpha \mathcal{F}_\mathcal{K} -\frac{6 \alpha^2 \mathcal{F}_\mathcal{KK} \mathcal{H}^2}{a^2 M^2} \right] \mathcal{H}, \\
C &= \left[ c_{123} \mathcal{F}_\mathcal{K} +\frac{2\alpha^2 \mathcal{H}^2 \mathcal{F}_\mathcal{KK}}{a^2 M^2} \right], \\
D &= c_{14}\mathcal{F}_\mathcal{K} k^2 \Psi - 3\alpha \mathcal{H}\left(\mathcal{F}_\mathcal{K} + \frac{6\alpha \mathcal{F}_\mathcal{KK}\mathcal{H}^2}{a^2 M^2} \right) (\mathcal{H}\Psi + \Phi'), \\
E &= \alpha\left( \mathcal{F}_\mathcal{K} + \frac{6\alpha \mathcal{H}^2 \mathcal{F}_\mathcal{KK}}{a^2 M^2} \right)(\mathcal{H}\Psi +\Phi').
\end{align} In \cite{EOS3} the $ABC$ matrix in \eqref{activation aether} was dubbed the activation matrix, as it determines which degrees of freedom are present, or activated, in the perturbed fluid variables. Inverting this then yields expressions for $V$ and $V'$ in terms of $\delta \rho$, $\theta^S$, the metric perturbations, $\Psi$ and $\Phi$, and their derivatives. Eliminating for these in $\Pi^S$ \eqref{CN Pi}, we find that we can write \begin{equation} \label{general Pi}
w_{\mathrm{de}}\Pi^S = A_1 \delta + A_2 (1+w) \theta^S + A_3 k^2 \Psi + A_4 (\mathcal{H}\Psi + \Phi'),
\end{equation} where \begin{align}
A_1 &= \frac{c_{13}}{c_{14}}, \\
A_2 &= \frac{3c_{13}\mathcal{H}}{3c_{123}+2\alpha \gamma_2}\left[1+ \frac{2(\mathcal{H}'-\mathcal{H}^2)}{\mathcal{H}^2}\gamma_2  + \frac{\alpha (1+2\gamma_2)}{c_{14}}  \right],  \\
A_3 &= \frac{2c_{13}\gamma_1}{3\alpha \mathcal{H}^2 \left( 2\gamma_1-1 \right) }, \\
A_4 &= \frac{2c_{13}\gamma_1(1+2\gamma_2)}{\mathcal{H}\left(2\gamma_1-1 \right)(3c_{123}+2\alpha \gamma_2)} \left[2 \left( \frac{c_{13}}{c_{14}} - \frac{(\mathcal{H}' - \mathcal{H}^2)}{\mathcal{H}^2}\gamma_2\right)  -1  \right]
\end{align} and we define the dimensionless functions \begin{equation}
\gamma_1 = \frac{\mathcal{K}\mathcal{F}_\mathcal{K}}{\mathcal{F}}, \quad
\gamma_2  = \frac{\mathcal{K}\mathcal{F}_\mathcal{KK}}{\mathcal{F}_\mathcal{K}}, \quad
\gamma_3 = \frac{\mathcal{K}\mathcal{F}_\mathcal{KKK}}{\mathcal{F}_\mathcal{KK}}.
\end{equation}

In the parlance of \cite{fRDE}, we write \eqref{general Pi} in terms of a set of dimensionless variables given in \autoref{table:GInv Var}, where $h_{\parallel} = h +6\eta$, $K = k/\mathcal{H}$, and $\epsilon_H=1-\mathcal{H}'/\mathcal{H}^2$.
\begin{table}
\begin{center}
	{\renewcommand{\arraystretch}{1.6}\begin{tabular}{ c | c | c }
		\textbf{Variable} &\textbf{Conformal Newtonian} & \textbf{Synchronous} \\ \hline
		$T$ & $\frac{h_{\parallel}'}{2\mathcal{H}K^2}$ & $0$\\
		$W$ & $\frac{1}{\mathcal{H}}X'-\epsilon_H(X+Y)$ & $\frac{1}{\mathcal{H}}X'-\epsilon_H(X+Y)$\\ 
		$X$ & $\frac{1}{\mathcal{H}}Z'+Y$ & $\frac{1}{\mathcal{H}}Z'+Y$  \\
		$Y$ & $\Psi$ & $\frac{1}{\mathcal{H}}T'+\epsilon_HT$\\
		$Z$ & $\Phi$ & $\eta-T$\\ 
		$\Delta$ & $\delta + 3\mathcal{H}(1+w)\theta^S$ & $\delta + 3\mathcal{H}(1+w)\theta^S$ \\
		$\hat{\Theta}$&$3\mathcal{H}(1+w)\theta^S$& $3\mathcal{H}(1+w)\theta^S +3(1+w)T$ \\
		$\delta\hat{P}$& $\delta P$ & $\delta P +P'T$
	\end{tabular}}
\end{center} \caption{Combinations of the metric perturbations and perturbed fluid variables are now written in terms of the dimensionless variables given in this table, in both the conformal Newtonian and synchronous gauges.} \label{table:GInv Var}
\end{table} Note that these new variables are gauge invariant except $T$, which we be important in the synchronous gauge. From this we can write \eqref{general Pi} as \begin{equation} \label{Pi}
w_{\mathrm{de}}\Pi^S = c_{\Pi \Delta} \Delta + c_{\Pi \Theta}\hat{\Theta} + c_{\Pi X}X + c_{\Pi Y}K^2 Y,
\end{equation} where \begin{align} \label{c_PiD}
c_{\Pi \Delta} &= \frac{c_{13}}{c_{14}}, \\
c_{\Pi \Theta} &= \frac{c_{13}}{3c_{123}+2\alpha \gamma_2}\left[1- 2\left( \epsilon_H \gamma_2+ \frac{c_{13}}{c_{14}}\right) \right] ,   \\
c_{\Pi X}&= \frac{2c_{13}\gamma_1(1+2\gamma_2)}{\left( 2\gamma_1-1 \right) (3c_{123}+2\alpha \gamma_2)}\left[ 2\left( \frac{c_{13}}{c_{14}} +\epsilon_H\gamma_2\right) -1 \right],  \\ \label{c_PiY}
c_{\Pi Y} &= \frac{2c_{13}\gamma_1}{3\alpha \left(1-2\gamma_1 \right) }.
\end{align}

In a similar fashion, we can eliminate $V$ and $V'$ in $\delta P$ and hence write the entropy perturbation as \begin{equation} \label{Gamma}
w_{\mathrm{de}}\Gamma = c_{\Gamma\Delta} \Delta + c_{\Gamma\Theta}\hat{\Theta} + c_{\Gamma W} W + c_{\Gamma X} X + c_{\Gamma Y} K^2 Y,
\end{equation} where \begin{align} \label{c_GammaD}
c_{\Gamma\Delta} &= \frac{\alpha (1+2\gamma_2)}{3c_{14}} - \frac{dP}{d\rho}, \\
c_{\Gamma\Theta} &= \frac{\alpha}{3 (3c_{123}+2\alpha \gamma_2)}\left[\left( 1-\frac{2c_{13}}{c_{14}}\right)(1+2\gamma_2)  - 6\epsilon_H \gamma_2\left( 1+\frac{2}{3}\gamma_3\right)   \right] + \frac{dP}{d\rho}, \\
c_{\Gamma W} &= \frac{2\gamma_1(1+2\gamma_2)}{3\left( 2\gamma_1-1 \right) }, \\
c_{\Gamma X} & = \frac{4\alpha\gamma_1}{3\left( 2\gamma_1-1 \right)(3c_{123}+2\alpha \gamma_2)}\left[\left(1+ \frac{c_{13}}{c_{14}}\right) (1+2\gamma_2)^2 + \frac{3c_{13}}{\alpha}\left(1+2\gamma_2\left[1-\epsilon_H\left(1+\frac{2}{3}\gamma_3 \right)  \right] \right) \right], \\ \label{c_GammaY}
c_{\Gamma Y} &= \frac{2\gamma_1(1+2\gamma_2)}{9\left( 1-2\gamma_1\right) }.
\end{align} Note that in \eqref{Pi} and \eqref{Gamma} the perturbed fluid variables are those for the dark energy fluid.

In order to ensure these results are truly gauge invariant, we must do the same calculation in the synchronous gauge. However, as mentioned previously, we now have an extra variable, $T$, to deal with. Therefore, let us suppose that in the synchronous gauge we find that \begin{align}
w_{\mathrm{de}}\Pi^S &= c_{\Pi \Delta} \Delta + c_{\Pi \Theta}\hat{\Theta} + c_{\Pi X}X + c_{\Pi Y}K^2 Y+c_{\Pi T}T, \\
w_{\mathrm{de}}\Gamma &= c_{\Gamma\Delta} \Delta + c_{\Gamma\Theta}\hat{\Theta} + c_{\Gamma W} W + c_{\Gamma X} X + c_{\Gamma Y} K^2 Y + c_{\Gamma T}T,
\end{align} with $c_{\Pi T}, c_{\Gamma T}\not=0$. If this was the case, $\Pi^S$ and $\Gamma$ would not be gauge invariant due to the presence of $T$ and so it must be that $c_{\Pi T}= c_{\Gamma T}=0$. Note that this was not necessary in the conformal Newtonian gauge as $T=0$ from \autoref{table:GInv Var}. We also require that in both gauges the coefficients are identical i.e. $c^{\mathrm{CN}}_{\Pi,\Gamma}=c^{\mathrm{Sync}}_{\Pi,\Gamma}$, because $\Delta, \hat{\Theta},W,X,$ and $Y$ are gauge invariant. Indeed, doing this calculation in the synchronous gauge we find that this is the case, and hence \eqref{Pi} and \eqref{Gamma} constitute the gauge invariant equations of state for the perturbations and are both presented simultaneously in the conformal Newtonian and synchronous gauges via \autoref{table:GInv Var}. For details of this calculation in the synchronous gauge see Appendix \ref{app:A}.

To ensure that no coefficient diverges we require that $\alpha$, $c_{14}$, $\gamma_1$, $2\gamma_1-1$, and $3c_{123}+2\alpha \gamma_2$ do not equal zero. If $\alpha=0$ then $\mathcal{K}=0$, removing the dynamics of Einstein-Aether completely, and so this must be excluded. As we will see later, to prevent a diverging sound speed for perturbations we must have that $c_{14}\not=0$ from \eqref{cs2}. The solution for $\gamma_1=0$ is constant $\mathcal{F}$, which is just the case of a cosmological constant with no Einstein-Aether and therefore has no perturbations, while setting $2\gamma_1-1=0$ yields $\rho_{\mathrm{m}}=0$ from the Friedmann equation. The case for disallowing $3c_{123}+2\alpha \gamma_2=0$ is more subtle. If this was true it would set the coefficient of $k^2V$ in \eqref{ThetaS_Rearranged} to zero and hence the activation matrix would be singular, i.e. we would be unable to eliminate the degrees of freedom $V$ and $V'$ from our equations using $\theta^S$. However, we note that this is not a strict condition and could in principle be true for some models as there is nothing that physically prevents this. For the designer $\mathcal{F}(\mathcal{K})$ in \eqref{Analytical F} this is non-zero and so all the $c_{\Pi,\Gamma}$ coefficients are well behaved.

Additionally, we can eliminate the metric perturbations in favour of the perturbed fluid variables for matter and dark energy as done in \cite{Horndeski2} for the Horndeski theory. This will allow us to write \eqref{Pi} and \eqref{Gamma} as \begin{align} \label{Pi2}
w_\mathrm{de}\Pi^S_\mathrm{de} &= c_{\Pi\Delta_\mathrm{de}}\Delta_\mathrm{de} + c_{\Pi\Theta_\mathrm{de}}\hat{\Theta}_\mathrm{de}+c_{\Pi\Delta_\mathrm{m}}\Delta_\mathrm{m} + c_{\Pi\Theta_\mathrm{m}}\hat{\Theta}_\mathrm{m}+c_{\Pi\Pi_\mathrm{m}}\Pi_{\mathrm{m}}^S, \\ \label{Gamma2}
w_\mathrm{de} \Gamma_{\mathrm{de}} &= c_{\Gamma\Delta_\mathrm{de}}\Delta_\mathrm{de} + c_{\Gamma\Theta_\mathrm{de}}\hat{\Theta}_\mathrm{de} +c_{\Gamma\Delta_\mathrm{m}}\Delta_\mathrm{m} + c_{\Gamma\Theta_\mathrm{m}}\hat{\Theta}_\mathrm{m}+ c_{\Gamma \Gamma_{\mathrm{m}}}\Gamma_{\mathrm{m}},
\end{align} where we now make explicit distinction between the perturbed fluid variables for matter and dark energy. In the notation of \autoref{table:GInv Var}, the perturbed Einstein equations take the form \cite{fRDE} \begin{align}
2W &= \Omega_\mathrm{m} \left(\frac{3 \delta \hat{P}_\mathrm{m}}{\rho_\mathrm{m}} + 2 w_\mathrm{m}\Pi^S_\mathrm{m} - 3 \hat{\Theta}_\mathrm{m} \right)  + \Omega_\mathrm{de} \left(\frac{3 \delta \hat{P}_\mathrm{de}}{\rho_\mathrm{de}} + 2 w_\mathrm{de}\Pi^S_\mathrm{de} - 3 \hat{\Theta}_\mathrm{de} \right), \\ \label{XEE}
2X & = \Omega_\mathrm{m} \hat{\Theta}_\mathrm{m} + \Omega_{\mathrm{de}} \hat{\Theta}_{\mathrm{de}}, \\ \label{YEE}
-\frac{2}{3}K^2 Y & =  \Omega_\mathrm{m} (\Delta_\mathrm{m} - 2w_\mathrm{m}\Pi^S_\mathrm{m}) + \Omega_\mathrm{de} (\Delta_\mathrm{de} - 2w_\mathrm{de}\Pi^S_\mathrm{de}), \\ \label{ZEE}
-\frac{2}{3}K^2 Z &= \Omega_\mathrm{m} \Delta_\mathrm{m} + \Omega_\mathrm{de} \Delta_\mathrm{de}.
\end{align} Substituting for these in \eqref{Pi} yields \begin{align} \label{Pi DM}
(1-3c_{\Pi Y}\Omega_{\mathrm{de}})w_\mathrm{de}\Pi^S_\mathrm{de}= &\left( c_{\Pi\Delta}-\frac{3}{2}c_{\Pi Y}\Omega_\mathrm{de}\right) \Delta_\mathrm{de}+\left(c_{\Pi\Theta}+\frac{1}{2}c_{\Pi X}\Omega_\mathrm{de} \right) \hat{\Theta}_\mathrm{de} \nonumber\\
&-\frac{3}{2}c_{\Pi Y}\Omega_\mathrm{m}\Delta_\mathrm{m}+\frac{1}{2}c_{\Pi X}\Omega_\mathrm{m}\hat{\Theta}_\mathrm{m}+3c_{\Pi Y}\Omega_\mathrm{m}w_\mathrm{m}\Pi^S_\mathrm{m},
\end{align} Similarly, the entropy perturbation becomes \begin{align} \label{Gamma DM}
\left( 1-\frac{3}{2}c_{\Gamma W}\Omega_\mathrm{de} \right)w_\mathrm{de}\Gamma_\mathrm{de} = &\left( c_{\Gamma \Delta}+\frac{3}{2}c_{\Gamma W}\Omega_\mathrm{de}\left. \frac{dP}{d\rho}\right|_\mathrm{de}-\frac{3}{2}c_{\Gamma Y}\Omega_\mathrm{de} \right) \Delta_\mathrm{de} +\frac{3}{2}\Omega_\mathrm{m}\left(c_{\Gamma W} \left. \frac{dP}{d\rho}\right|_\mathrm{m} -c_{\Gamma Y}\right)\Delta_\mathrm{m} \nonumber \\ &+\left[ c_{\Gamma \Theta}-\frac{3}{2}c_{\Gamma W}\Omega_\mathrm{de}\left( 1 + \left. \frac{dP}{d\rho}\right|_\mathrm{de} \right)+\frac{1}{2}c_{\Gamma X}\Omega_\mathrm{de}  \right]\hat{\Theta}_\mathrm{de} \nonumber \\
&+\frac{1}{2}\left[c_{\Gamma X}-3c_{\Gamma W} \left( 1 + \left. \frac{dP}{d\rho}\right|_\mathrm{m} \right)  \right]  \Omega_\mathrm{m}\hat{\Theta}_\mathrm{m} +\frac{3}{2}c_{\Gamma W}\Omega_\mathrm{m}w_\mathrm{m}\Gamma_\mathrm{m}.
\end{align} Note that \eqref{Pi DM} and \eqref{Gamma DM} are completely general and not specific to Generalized Einstein-Aether. If for any theory $w_{\mathrm{de}}\Pi^S$ and $w_{\mathrm{de}}\Gamma$ can be written as \eqref{Pi} and \eqref{Gamma}, then \eqref{Pi DM} and \eqref{Gamma DM} will also be true automatically.

From these expressions we can derive the sound speed for scalar perturbations. Starting from the perturbed conservation equations, \eqref{ConsEq1} and \eqref{ConsEq2}, we can deduce that \begin{equation}
\delta''+\cdots+k^2c_s^2\delta=F(h,\eta,...).
\end{equation} Therefore, extracting the coefficient of $k^2 \delta$ we find that \begin{equation} \label{cs2}
c_{\mathrm{s}}^2=\frac{1}{c_{14}}\left(c_{123}+\frac{2}{3}\alpha\gamma_2 \right).
\end{equation} In general, the sound speed of scalar perturbations varies with time due to $\mathcal{F}$. To ensure subluminal propagation and stable growth of perturbations, we require that $0\leq \frac{1}{c_{14}}\left(c_{123}+\frac{2}{3}\alpha\gamma_2 \right) \leq 1$.

From here, we could attempt to obtain constraints on the $\left\lbrace c_i\right\rbrace $ coefficients by appealing to the behaviour of perturbations in the limit of Minkowski space, as in \cite{Lim}. However, as we have directly coupled the evolution of $\mathcal{F}$ to $a(t)$ via a designer approach, we argue that no sensible Minkowski limit exists for this theory once this connection has been made. For a brief discussion of this see Appendix \ref{app:B}. In the context of the Equation of State approach, in the limit of $H \rightarrow 0$ we see that $\rho$, $P \rightarrow 0$ from \eqref{rho} and \eqref{Pressure}. Therefore, the expressions for $w_{\mathrm{de}}\Pi^S$ and $w_{\mathrm{de}}\Gamma$ cannot be computed since $w_{\mathrm{de}}\Pi^S$ appears as $P\Pi^S$ from the perturbed energy momentum tensor  \eqref{ij projection} and $w_{\mathrm{de}}\Gamma$ can be written as $w_{\mathrm{de}}\rho\Gamma = \left(\frac{\delta P}{\delta \rho}-\frac{dP}{d\rho} \right) \delta \rho$.
\subsection{Special cases}

\subsubsection{$w_{\mathrm{de}}=-1$}
Consider the case where we have exactly $w_{\mathrm{de}}=-1$, equivalent to $\Lambda\mathrm{CDM}$. From \autoref{sect:Background} we have an analytical solution given by \eqref{Analytical F} and in this case the $c_\Pi$ and $c_\Gamma$ coefficients reduce to \begin{equation}
c_{\Pi\Delta} = \frac{c_{13}}{c_{14}}, \quad
c_{\Pi\Theta} = \frac{1}{2}\left(1+\epsilon_H \right) - \frac{c_{13}}{c_{14}},\quad
c_{\Pi X} = 0, \quad
c_{\Pi Y} =  -\frac{c_{13}}{3\alpha}\left( 1+\frac{M^2\mathcal{F}_0}{6\Omega_{\mathrm{de},0}H_0^2}\right)\left( \frac{H}{H_0}\right)   , 
\end{equation} and also \begin{equation}
c_{\Gamma \Delta} = -c_{\Gamma \Theta} = -\frac{dP}{d\rho}=1, \quad
c_{\Gamma W} = c_{\Gamma X} = c_{\Gamma Y} = 0,
\end{equation} and hence $\Gamma = \delta$.  Here we see that from $c_{\Pi Y}$, as with the background evolution, $M$ and $\mathcal{F}_0$ are degenerate.

This case is indistinguishable from $\Lambda\mathrm{CDM}$ at background order, but at the level of linear perturbations they are not the same. Therefore, geometrical cosmological tests such as SNe and BAOs would not be able to observe a difference between $\Lambda\mathrm{CDM}$ and Generalized Einstein-Aether with $w_{\mathrm{de}}=-1$, whereas probes which are sensitive to perturbations, such as weak lensing, will be different and can in principle distinguish between them.

From \eqref{Analytical F} we note that the $\Lambda$CDM limit is when $\mathcal{F}_0 = -6H_0^2\Omega_{\mathrm{de},0}/M^2$ and so $\mathcal{F} = -6H_0^2\Omega_{\mathrm{de},0}/M^2$ \eqref{Analytical F}. This case corresponds to the cosmological constant in the Friedmann equation. Indeed, this also is reflected at the level of linear perturbations since $\mathcal{F}_\mathcal{K} = 0$ and so all the perturbed fluid variables and the equation of motion for $V$ in \autoref{sect:Pert} are zero, as in $\Lambda$CDM. However, it seems that there is a discontinuity in taking the limit of $\mathcal{F}_0 \rightarrow -6H_0^2\Omega_{\mathrm{de},0}/M^2$, since in this limit the $c_{\Pi,\Gamma}$ coefficients become \begin{equation}
c_{\Pi\Delta} = \frac{c_{13}}{c_{14}}, \quad c_{\Pi\Theta} = \frac{1}{2}(1+\epsilon_H)-\frac{c_{13}}{c_{14}}, \quad  c_{\Pi X} = c_{\Pi Y} = 0
\end{equation} and \begin{equation}
c_{\Gamma\Delta} = -c_{\Gamma\Theta} = 1, \quad c_{\Gamma W} = c_{\Gamma X} = c_{\Gamma Y} = 0
\end{equation} i.e. $\Pi^S$ and $\Gamma$ are non-zero in this limit, but are zero if $\mathcal{F}_0 = -6H_0^2\Omega_{\mathrm{de},0}M^2$ exactly. This is a property shared by $f(R)$ models in the limit of $B_0 \rightarrow 0$.

\subsubsection{Power law}
For a general power law with $\mathcal{F} \propto (\pm\mathcal{K})^n$ as studied in \cite{Ferreira1,Ferreira2,Ferreira3}, the coefficients become \begin{align}
c_{\Pi\Delta} &= \frac{c_{13}}{c_{14}}, \\
c_{\Pi\Theta} &= \frac{c_{13}}{(2n+1)\alpha -6c_2}\left[1-2\left( \epsilon_H(n-1)-\frac{c_{13}}{c_{14}}\right)  \right], \\
c_{\Pi X} &= \frac{2nc_{13}}{(2n+1)\alpha -6c_2 }\left[\frac{2c_{13}}{c_{14}} - 1 + 2\epsilon_H (n-1) \right],  \\
c_{\Pi Y} &=  \frac{2nc_{13}}{3\alpha(1-2n)},
\end{align} and \begin{align}
c_{\Gamma \Delta} &= \frac{(2n-1)\alpha}{3c_{14}}-\frac{dP}{d\rho}, \\
c_{\Gamma \Theta} &= \frac{(2n-1)\alpha}{3\left[ (2n+1)\alpha -6c_2\right] }\left[ 1-2\epsilon_H(n-1)-\frac{c_{13}}{c_{14}} \right] +\frac{dP}{d\rho}, \\
c_{\Gamma W} &= \frac{2}{3}n, \\
c_{\Gamma X} &= \frac{4n}{3\left[ (2n+1)\alpha -6c_2 \right] }\left[ \frac{\alpha(2n-1)(c_{13}+c_{14})}{c_{14}}+3c_{13}\left( 1 - \frac{2}{3}\epsilon_H(n-1) \right) \right], \\
c_{\Gamma Y} &= -\frac{2}{9}n.
\end{align} Note that $c_{\Pi Y}$ is singular for the case of $n= \frac{1}{2}$. Although $\mathcal{F} \propto (\pm \mathcal{K})^{1/2}$ is also a solution to \eqref{Background DE}, inserting this into the Friedmann equation \eqref{Modified Friedmann} shows that this case corresponds to an absence of dark energy at the level of background cosmology.

\subsection{Dynamics of linear perturbations in the scalar sector} %change introduction

The dynamics of scalar perturbations can be computed via the perturbed fluid equations in \eqref{ConsEq1} and \eqref{ConsEq2}. We will use the designer $\mathcal{F}(\mathcal{K})$ model via \eqref{Background DE}. Following the notation of \autoref{table:GInv Var} we rewrite these equations as \begin{equation} \label{ConsEq1GI}
\dot{\Delta}-3w\Delta+g_{\mathrm{K}}\epsilon_H \hat{\Theta}-2w\Pi^S = 3(1+w) X,
\end{equation} \begin{equation} \label{ConsEq2GI}
\dot{\hat{\Theta}} + 3\left(\frac{dP}{d\rho} - w + \frac{1}{3} \epsilon_H \right)\hat{\Theta} - 3\frac{dP}{d\rho}\Delta - 2w\Pi^S - 3w\Gamma=3(1+w)Y,
\end{equation} where $g_{\mathrm{K}}=1+\frac{K^2}{3\epsilon_H}$ and, for this section only, over-dots denote differentiation with respect to the logarithmic scale factor, $\log a$. For a cold, pressureless matter fluid with $w_{\mathrm{m}}= \Pi^S_{\mathrm{m}} = \Gamma_{\mathrm{m}} = 0$ and assuming $w_\mathrm{de}$ constant, \eqref{ConsEq1GI} and \eqref{ConsEq2GI} yield 4 differential equations for the dark energy and matter perturbed fluid variables, given by \begin{align} \label{DeltaM}
\dot{\Delta}_{\mathrm{m}}+g_{\mathrm{K}}\epsilon_H \hat{\Theta}_{\mathrm{m}} &= 3X, \\
 \dot{\hat{\Theta}} _{\mathrm{m}}+\epsilon_H\hat{\Theta}_{\mathrm{m}} &=3Y, \\
\dot{\Delta}_{\mathrm{de}}-3w_\mathrm{de}\Delta_{\mathrm{de}}+g_{\mathrm{K}}\epsilon_H \hat{\Theta}_{\mathrm{de}}-2w_\mathrm{de}\Pi^S_{\mathrm{de}} &= 3(1+w_\mathrm{de}) X, \\ \label{ThetaDE}
 \dot{\hat{\Theta}}_{\mathrm{de}} + \epsilon_H \hat{\Theta}_{\mathrm{de}} - 3w_\mathrm{de}\Delta_{\mathrm{de}} - 2w_\mathrm{de}\Pi^S_{\mathrm{de}} - 3w_\mathrm{de}\Gamma_{\mathrm{de}}&=3(1+w_\mathrm{de})Y. 
 \end{align} With these, the dynamics of the Newtonian gravitational potential, $Z = \Phi$, can be computed directly from the perturbed Einstein equation in \eqref{ZEE} or via $\dot{Z}= X-Y$ from the definition of $Z$ in \autoref{table:GInv Var}. Note that in \autoref{table:GInv Var} the variables and derivatives are in conformal time, not the scale factor. To solve these equations we will opt to specify $\Pi^S_{\mathrm{de}}$ and $\Gamma_{\mathrm{de}}$ terms of the perturbed fluid variables for dark energy and matter, given in \eqref{Pi DM} and \eqref{Gamma DM}. The variables $X$ and $Y$ are also specified in terms of the perturbed fluid variables via the perturbed Einstein equations in \eqref{XEE} and \eqref{YEE}. We note that this is not the only way to proceed. For example, instead of the perturbed fluid variables we could have opted to work with the metric perturbation variables $W,X,Y$, and $Z$. For more details see \cite{fRDE}. We set the initial conditions as described in \cite{fRDE}. They are set at $z=100$ such that $\Delta_{\mathrm{de}} = \hat{\Theta}_\mathrm{de} = 0$, $\Omega_{\mathrm{m}}\Delta_{\mathrm{m}} = -\frac{2}{3} K^2 Z$, $\Omega_{\mathrm{m}}\hat{\Theta}_\mathrm{m}=2X$, and $X=Y=Z$. Since the behaviour of the perturbations will also depend of the specific choice of $\left\lbrace c_i \right\rbrace $ and not just $\alpha$, we will fix $c_1=1$, $c_2=1$, $c_3=1$, and $c_4=-3$. This choice is somewhat arbitrary, other than ensuring the subluminal propagation of the perturbations \eqref{cs2}.
 
 \begin{figure}
 	\centering
 	\includegraphics[width=0.497\textwidth]{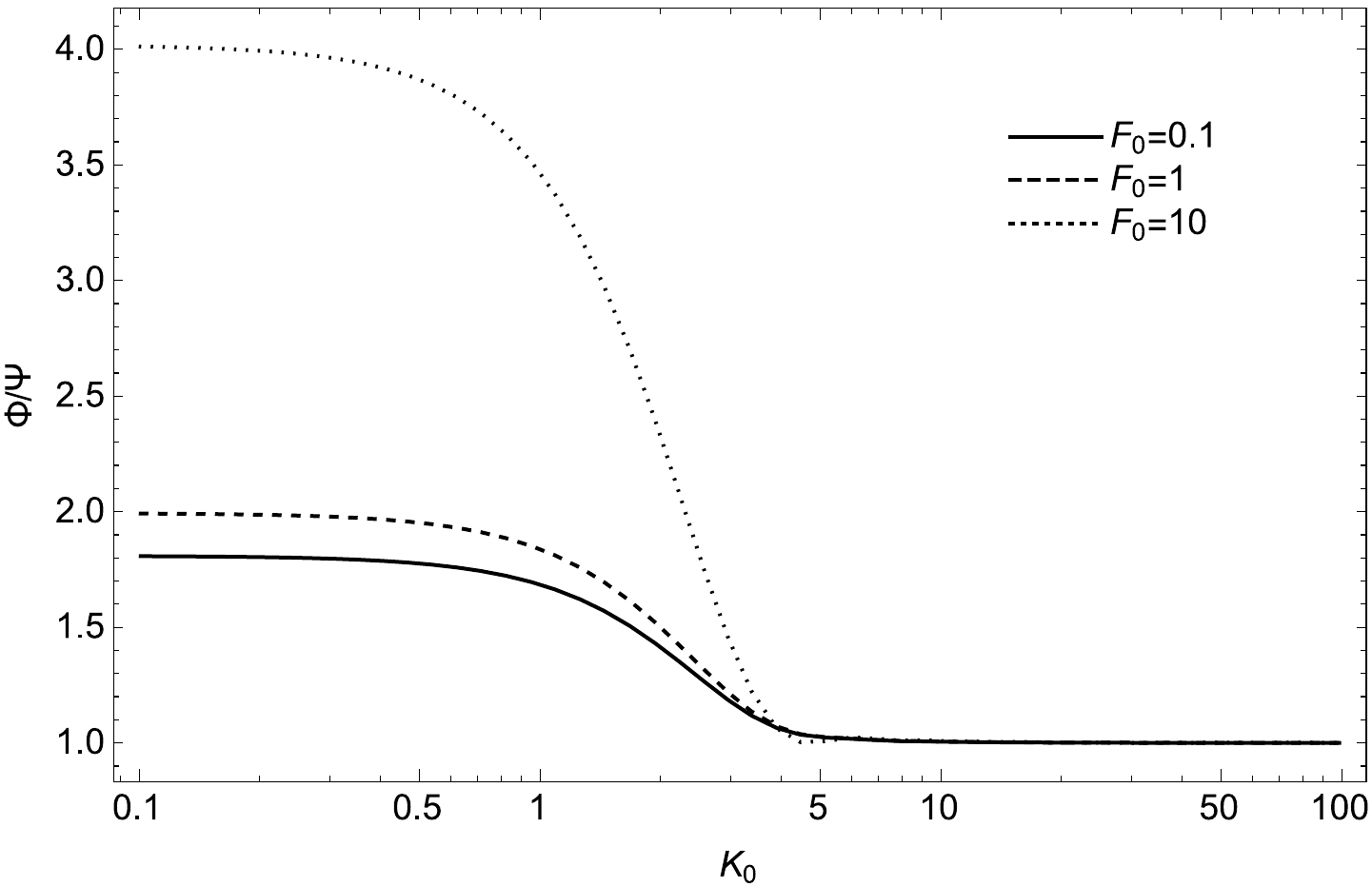}
 	\includegraphics[width=0.497\textwidth]{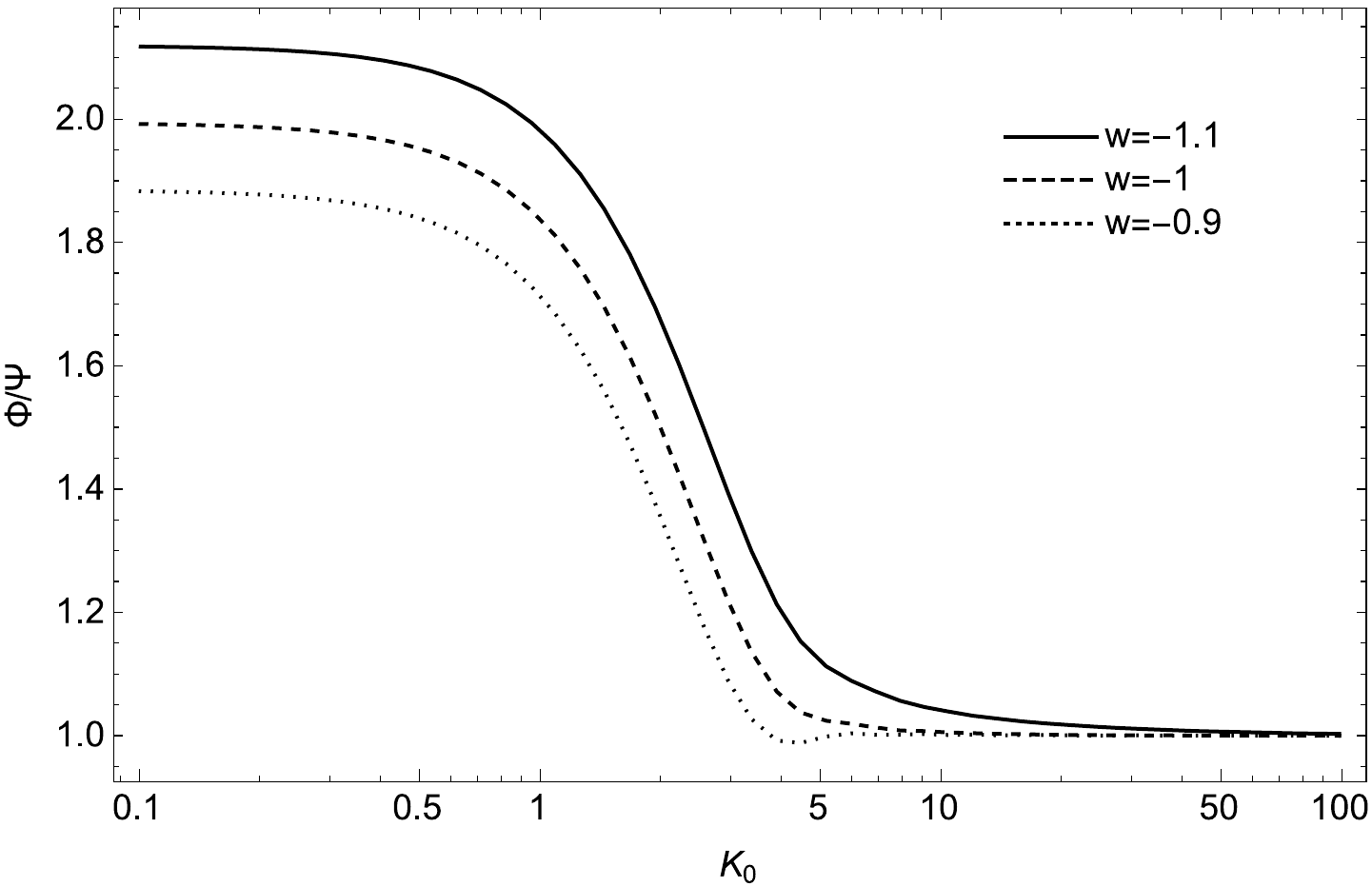}
 	\caption{\textit{Left panel}: The spectrum of $\Phi/\Psi$, or $Z/Y$, at $a=1$ as a function of scale for varying $\mathcal{F}_0$ and $w_\mathrm{de}=-1$.
 		\textit{Right panel}: The spectrum of $\Phi/\Psi$ at $a=1$ as a function of scale for a General Einstein-Aether fluid with $w_\mathrm{de}$ varying around $-1$ and $\mathcal{F}_0=1$.}
 	\label{fig:ZYk}
 \end{figure}

 \begin{figure}
	\centering
	\includegraphics[width=0.497\textwidth]{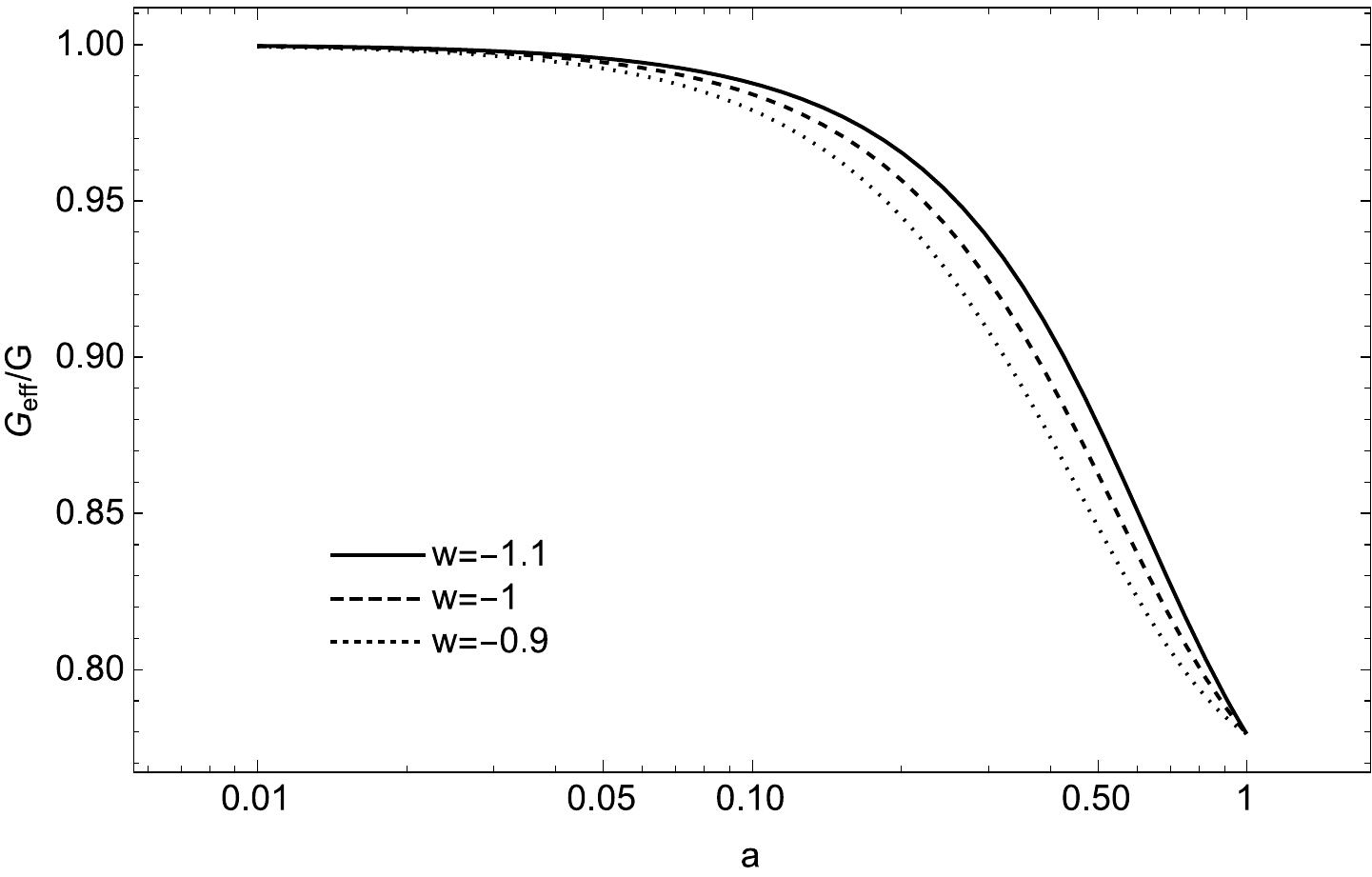}
	\caption{The evolution of the effective Newton's constant, $G_\mathrm{eff}/G$, is shown for varying $w_\mathrm{de}$ around $-1$.}
	\label{fig:Geff}
\end{figure}
 
 We investigate how the ratio of the Newtonian potentials vary with scale. From \autoref{fig:ZYk}, we see that at $a=1$, the large scale behaviour of $\Phi/\Psi$ is highly dependent on $\mathcal{F}_0$, while this is less so for $w_\mathrm{de}$ near $-1$. We see that $\Phi/\Psi$ tends to a constant in both the large and small $K_0$ regimes. In all cases the small scale behaviour is such that $\Phi = \Psi$ and so this indicates a vanishing $w_\mathrm{de}\Pi_{\mathrm{de}}^S$ for small scales. Note that $K_0 = 1$ corresponds to a scale of $3.35 \times 10^{-4} h \, \mathrm{Mpc}^{-1}$. 
 
 In the regime $K \gg 1$ we find that the $\left\lbrace \hat{\Theta}_i \right\rbrace $ are negligible and so we can write $w_{\mathrm{de}}\Pi^S_\mathrm{de} \approx c_{\Pi\Delta_{\mathrm{de}}}\Delta_{\mathrm{de}} + c_{\Pi\Delta_{\mathrm{m}}}\Delta_{\mathrm{m}}$. From equations \eqref{DeltaM} to \eqref{ThetaDE} we compute the second order differential equations for $\left\lbrace \Delta_i \right\rbrace $, given by \begin{align} \label{2ODE DeltaM}
\ddot{\Delta}_\mathrm{m} +(2-\epsilon_H)\dot{\Delta}_\mathrm{m} - \frac{3}{2}\Omega_{\mathrm{m}}\Delta_{\mathrm{m}}&=\frac{3}{2}\Omega_{\mathrm{de}}\Delta_{\mathrm{de}},\\ \label{2ODE DeltaDE}
 \ddot{\Delta}_\mathrm{de}+(5-\epsilon_H)\dot{\Delta}_{\mathrm{de}} +\frac{2}{3}c_{\Pi\Delta_{\mathrm{de}}}K^2\Delta_{\mathrm{de}}&=-\frac{2}{3}c_{\Pi\Delta_{\mathrm{m}}}K^2\Delta_{\mathrm{m}},
 \end{align} where we have also used the Einstein equations for $X$ \eqref{XEE} and $Y$ \eqref{YEE}, with $w_\mathrm{de}=-1$. Note that in \eqref{2ODE DeltaM} the secondary source term arising from $w_\mathrm{de}\Pi^S_\mathrm{de}$ is subdominant compared to $\Omega_{\mathrm{de}}\Delta_{\mathrm{de}}$ and so we have neglected this. From \eqref{XEE} and \eqref{YEE}, for small scales we have that \begin{equation} \label{Y/Z}
 \frac{Y}{Z} = 1-\frac{2\Omega_{\mathrm{de}}( c_{\Pi\Delta_{\mathrm{de}}}\Delta_{\mathrm{de}} +c_{\Pi\Delta_{\mathrm{m}}}\Delta_{\mathrm{m}})}{\Omega_{\mathrm{de}}\Delta_{\mathrm{de}}+\Omega_{\mathrm{m}}\Delta_{\mathrm{m}}},
 \end{equation} where the second term must be negligible from \autoref{fig:ZYk}. In order to explain this, note that from \eqref{2ODE DeltaDE} we must have that the solution tends to the particular solution \begin{equation} \label{Particular Solution}
 c_{\Pi\Delta_{\mathrm{de}}}\Delta_{\mathrm{de}} = -c_{\Pi\Delta_{\mathrm{m}}}\Delta_{\mathrm{m}}.
 \end{equation}  Hence, the second term in \eqref{Y/Z} is always negligible regardless of what the $\left\lbrace c_{\Pi\Delta_i}\right\rbrace $ are. Therefore, a vanishing anisotropic stress at small scales is a generic feature of these designer $\mathcal{F}(\mathcal{K})$ models.
 
Using \eqref{Particular Solution} in \eqref{2ODE DeltaM}, we find that this becomes the standard differential equation for the matter overdensity with Newton's constant replaced with an effective Newton's constant, $G_{\mathrm{eff}}$, given by \begin{equation}
 \frac{G_{\mathrm{eff}}}{G} = 1 - \frac{\Omega_{\mathrm{de}}c_{\Pi\Delta_{\mathrm{m}}}}{\Omega_{\mathrm{m}}c_{\Pi\Delta_{\mathrm{de}}}}
 \end{equation} and the evolution of this is shown in \autoref{fig:Geff}. We see that the ratio $G_\mathrm{eff}/G$ is always of order unity but that for our choice of $\left\lbrace c_i \right\rbrace $ it decreases to $G_\mathrm{eff} \approx 0.78 G$ at $a=1$, which should lead to a suppression of structure at late times compared to $\Lambda$CDM. We leave this as a matter for future investigation. We also observe that increasing $w_\mathrm{de}$ causes $G_\mathrm{eff}/G$ to decay faster at early times, while the opposite is true for decreasing $w_\mathrm{de}$. It is interesting to note that the value of $G_\mathrm{eff}/G$ for different $w_\mathrm{de}$ initially diverge and then converge again at $a=1$. Note that what we have called $G_{\mathrm{eff}}$ is different to that in \cite{Ferreira2}, for example, which is derived from the modified Poisson equation.
\begin{figure}
	\centering
	\includegraphics[width=0.497\textwidth]{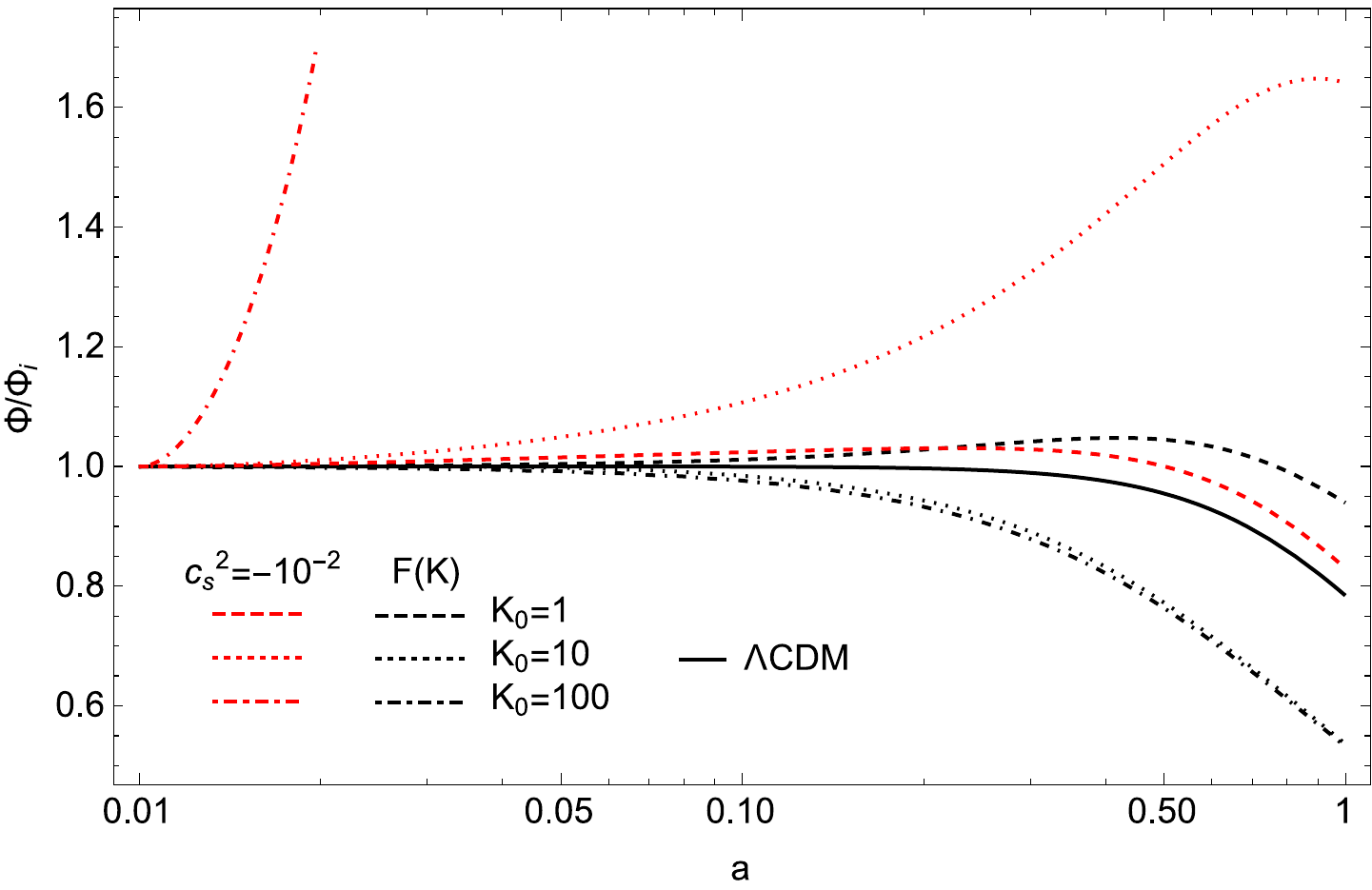}
	\includegraphics[width=0.497\textwidth]{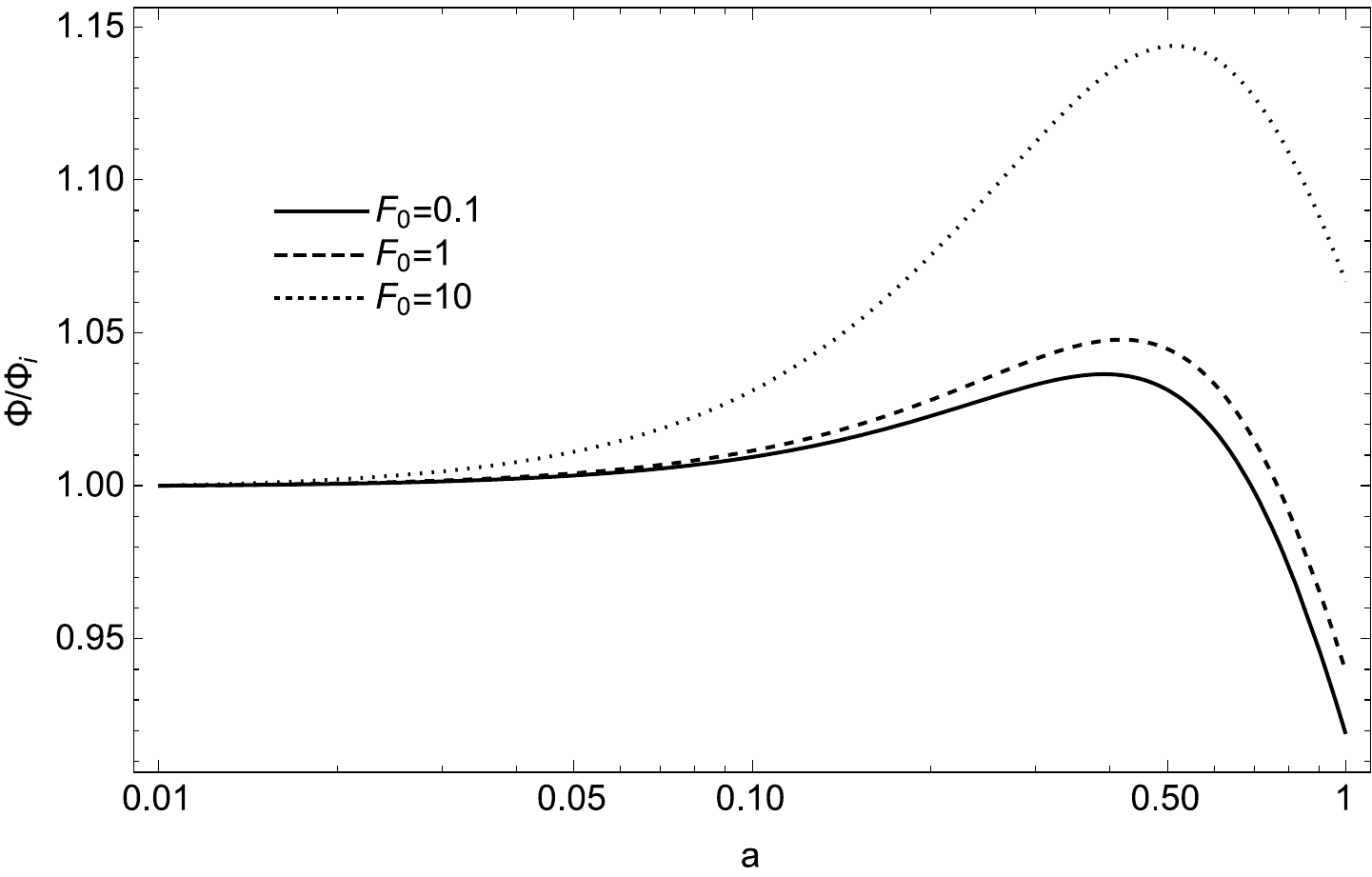}\\
	\includegraphics[width=0.497\textwidth]{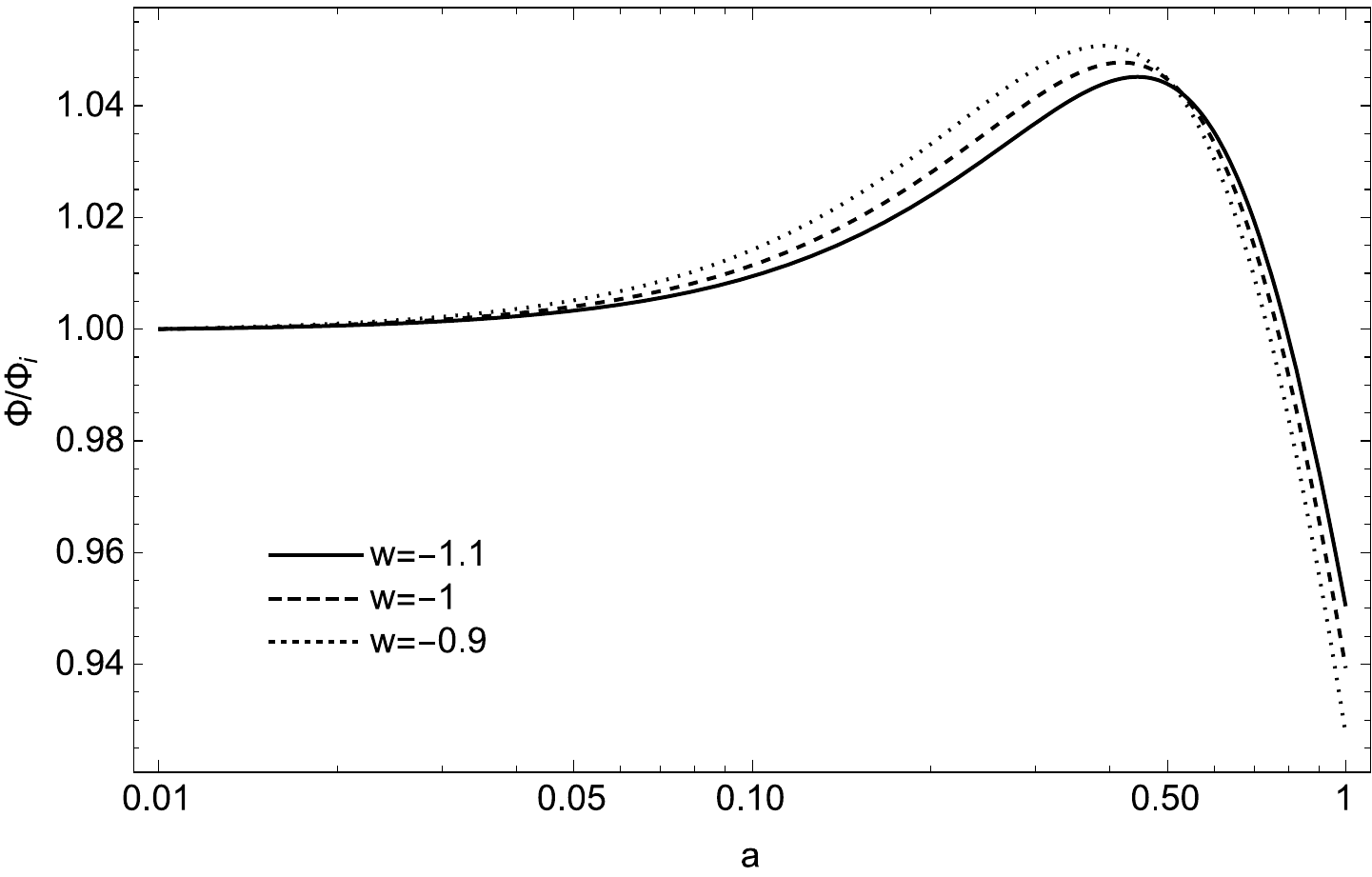}
	\caption{\textit{Top left panel}: The evolution of the Newtonian potential, $\Phi$, in $\Lambda$CDM (black solid line) and for different scales in a designer $\mathcal{F}(\mathcal{K})$ model (dashed and dotted lines) for $\mathcal{F}_0 = 1$ and $w_\mathrm{de}=-1$. Note that the potential for the $\Lambda$CDM model is scale independent. For comparison we also show the evolution of $\Phi$ with the presence of a dark energy fluid with $w_\mathrm{de}=-1$ and a constant negative squared sound speed of $c_\mathrm{s}^2 = -10^{-2}$ (red lines), calculated using \eqref{Phi ODE}.
	\textit{Top right panel}: The evolution of $\Phi$ in a Generalized Einstein-Aether universe with varying $\mathcal{F}_0$ for $w_\mathrm{de}=-1$ and $K_0=1$ fixed. \textit{Bottom panel}: The evolution of $\Phi$ for a General Einstein-Aether fluid with $w_\mathrm{de}$ varying around $-1$, with $\mathcal{F}_0=1$ and $K_0=1$ fixed.}
	\label{fig:Phi Potential a}
\end{figure}

We also investigate the evolution for the Newtonian potential, $\Phi$, as a function of $a$ and this is shown in \autoref{fig:Phi Potential a}. We see that for a designer $\mathcal{F}(\mathcal{K})$ model which mimics a $\Lambda$CDM background the evolution is now sensitive to the scale, where $K_0= k/H_0$, unlike the case of a cosmological constant. The amplitude of $\Phi$ grows with respect to $\Lambda$CDM for large scales, while for smaller scales the amplitude is suppressed. For scales $K_0 \lesssim 1$, we see that $\Phi$ initially grows before reaching a maximum and then decays due to the increasing contribution from dark energy. A similar feature was also observed in \cite{Ferreira2} for their power law model of $\mathcal{F}$. We note that this is very similar to other models which introduce a new cosmological fluid with a negative squared sound speed, $c_\mathrm{s}^2=\delta P/\delta \rho$. We solve the differential equation governing the evolution of $\Phi$ \cite{PhiODE1,PhiODE2} \begin{equation} \label{Phi ODE}
\frac{d^2\Phi}{da^2}+\left(\frac{1}{\mathcal{H}}\frac{d\mathcal{H}}{da}+\frac{4}{a}+3\frac{c_\mathrm{s}^2}{a} \right)\frac{d\Phi}{da} + \left[ \frac{2}{a\mathcal{H}} \frac{d\mathcal{H}}{da}+\frac{1}{a^2}(1+3c_\mathrm{s}^2)+\frac{c_\mathrm{s}^2k^2}{a^2\mathcal{H}^2}\right]\Phi = 0, 
\end{equation} provided there is zero anisotropic stress and so $\Phi = \Psi$. In models where $c_\mathrm{s}^2 < 0 $ we observe the same behaviour for $\Phi$ rising to a maximum before decaying, as seen in \autoref{fig:Phi Potential a}. In these models, the initial growth is due to an imaginary $c_\mathrm{s}^2$  causing an unstable growth of perturbations. However, as dark energy begins to dominate $\Phi$ decays as in $\Lambda$CDM. This feature is enhanced for smaller scales until the effect of dark energy in unable to overcome the unstable growth of perturbations and $\Phi$ grows exponentially as seen for $K_0 = 100$ in \autoref{fig:Phi Potential a}. While a fluid with $c_{\mathrm{s}}^2 < 0$ is unphysical, it is interesting to note that this feature appears in a designer $\mathcal{F}(\mathcal{K})$ universe without the need for $c_{\mathrm{s}}^2 < 0$. Indeed, the $\left\lbrace c_i \right\rbrace $ coefficients were chosen to avoid this. Moreover, we see that for designer $\mathcal{F}(\mathcal{K})$ the opposite occurs compared to $\Lambda$CDM and that as we go to smaller scales this feature is suppressed rather than enhanced.

\subsection{Vector and tensor sectors}
We can also calculate the equation of state for the vector sector. In this case, the function we specify is $\Pi^V=\Pi^V(\theta^V)$. Since we only have one function, $\theta^V$, to eliminate the vector degree of freedom, $B^V$, it may not seem possible as $\theta^V \equiv \theta^V(B^{V},B^{V'},B^{V''})$, as seen from \eqref{Theta V}. However, in a similar process to the scalar sector, we can use the perturbed equation of motion \eqref{PertVectorEoMVS} to eliminate derivatives of $B^V$. In doing so, \eqref{Theta V} becomes \begin{equation}
a^2\rho(1+w_\mathrm{de})\theta^V = \frac{1}{2}c_{13}\mathcal{F}_{\mathcal{K}}(kB^V-h^{V'}).
\end{equation} Inserting this into \eqref{PiV}, we obtain the equation of state for perturbations in the vector sector as \begin{equation} \label{PiVEoS}
w_{\mathrm{de}}\Pi_{\mathrm{de}}^V=\left[(1-3w_{\mathrm{de}})(1+w_{\mathrm{de}})\mathcal{H} \right]\theta_{\mathrm{de}}^V +(1+w_{\mathrm{de}})\theta_{\mathrm{de}}^{V'}.
\end{equation} Note that this is exactly the same as the perturbed conservation equation and is, therefore, a tautology. To proceed we use the vector Einstein equations, given by \begin{align} \label{Vector EFE Theta}
-\frac{1}{2a^2}h^{V'}&=8\pi G\rho_\mathrm{m}(1+w_\mathrm{m})\theta^V_\mathrm{m}+ \rho_{\mathrm{de}}(1+w_{\mathrm{de}})\theta^V_{\mathrm{de}}, \\
\frac{1}{6\mathcal{H}^2}h^{V''}+\frac{1}{3\mathcal{H}}h^{V'}&=\Omega_\mathrm{m} w_\mathrm{m} \Pi_\mathrm{m}+\Omega_\mathrm{de} w_\mathrm{de} \Pi_\mathrm{de}.
\end{align} Differentiating \eqref{Vector EFE Theta} and eliminating for $\theta^{V'}_{\mathrm{de}}$ and the metric perturbations in \eqref{PiVEoS}, we find that \begin{equation} \label{Pi V}
w_{\mathrm{de}}\Pi^V_{\mathrm{de}} = \mathcal{H}(1+w_{\mathrm{de}})\theta^V_{\mathrm{de}}+\frac{\Omega_{\mathrm{m}}}{\Omega_{\mathrm{de}}}\left[ \mathcal{H}(1+w_{\mathrm{m}})\theta^V_{\mathrm{m}}-w_{\mathrm{m}}\Pi^V_{\mathrm{m}} \right].
\end{equation} 

For the tensor sector, since there are no new tensor degrees of freedom, $\Pi^T$ can only be a function of $h^T$ and its derivatives. Therefore, \eqref{h plus} immediately constitutes the equation of state for tensor perturbations and is given by \begin{align} \label{Pi T}
3 \alpha \mathcal{H}^2 \left( \mathcal{F}_{\mathcal{K}} - \frac{\mathcal{F}}{2\mathcal{K}}\right) w_{\mathrm{de}}\Pi_{\mathrm{de}}^T = &-c_{13} \left(\mathcal{F}_{\mathcal{K}} \mathcal{H}+\frac{1}{2}\mathcal{F}_{\mathcal{KK}}\mathcal{K}'\right)h^{T'} -\frac{1}{2}c_{13} \mathcal{F}_{\mathcal{K}} h^{T''}.
\end{align} We can, therefore, derive the modification to the propagation speed of gravitational waves, due to the presence of the Aether field. Projecting out the tensor mode of the $ij$-component of the Einstein equation \eqref{Einstein Equation} yields
%\begin{equation}
%2a^2 \delta G^i _j &= -\left( h'' +2 \mathcal{H}h' - \nabla^2 h +\partial_k \partial_l h^{kl} \right) \delta^i _j + h''^i \hspace{0.1mm}_j +2 \mathcal{H}h'^i \hspace{0.1mm}_j - \nabla^2 h^i \hspace{0.1mm}_j +\partial^i \partial_k h^k \hspace{0.1mm}_j +\partial_j \partial_k h^{ik} -\partial^i \partial_j h
%\end{equation}
\begin{equation}
a^2 (\hat{l}_i\hat{l}^j - \hat{m}_i \hat{m}^j)\delta G^i _j = h^{T''}  +2\mathcal{H}h^{T'}  + k^2h^T = a^2 (\hat{l}_i\hat{l}^j - \hat{m}_i \hat{m}^j)\delta U^i _j = 2a^2 P\Pi^T,
\end{equation} assuming that the matter energy-momentum tensor contributes zero anisotropic stress. Hence, from \eqref{Pi T} we find that \begin{equation}
(1 + c_{13} \mathcal{F}_{\mathcal{K}} ) h^{T''} + 2\left[ \mathcal{H} + c_{13} \left(  \mathcal{F}_\mathcal{K} \mathcal{H} +\frac{1}{2}\mathcal{F}_{\mathcal{KK}}\mathcal{K}' \right)   \right] h^{T'} + k^2 h^T = 0
\end{equation} and so gravitational waves propagate with speed \begin{equation}
c_{\mathrm{grav}}^2 = \displaystyle \frac{1}{1+c_{13}\mathcal{F}_\mathcal{K}}.
\end{equation} We see that, in general, the propagation speed of gravitational waves is time dependent via $\mathcal{F}$. This is consistent with the result in \cite{Ferreira2}. It is often argued that on the grounds of causality that we should constrain $c_{\mathrm{grav}} \leq 1$, as was said for the scalar perturbations \eqref{cs2}. Indeed, this is the standard argument that was often made in previous work, for example see \cite{Lim} and Appendix \ref{app:B}. However, if gravitational waves were to propagate subluminally we would expect the existence of gravitational Cherenkov radiation, of which very stringent constraints have been placed \cite{Cherenkov}. See also \cite{Ferreira1} for a discussion. It was also noted in \cite{Cherenkov} that the constraint for $c_{\mathrm{grav}} \geq 1$ were much weaker. Moreover, given that this is already a Lorentz violating theory it could be argued that $c_{\mathrm{grav}} \geq 1$ may not be a problem, however we do not discuss this further.

\section{Discussion and conclusions} \label{sect:Conc}

In this paper the background dynamics of Generalized Einstein-Aether are studied using a designer approach. We find that only one form of $\mathcal{F}$ gives rise to a fluid species with $w_\mathrm{de}=-1$ exactly \eqref{CCEA} for a `designer' $\mathcal{F}(\mathcal{K})$ model. However, we see that at the level of linear perturbations this model is not the same as $\Lambda$CDM. We obtain a differential equation for general values of constant $w_\mathrm{de}$ \eqref{Background DE}, which is solved numerically to see how this model behaves as we vary the parameters in the theory, shown in \autoref{fig:F Evolution}. We also find that the background evolution is independent of the choice of $\left\lbrace c_i \right\rbrace $. For $w_\mathrm{de}=-1$ there is an analytical solution for $\mathcal{F}$ given by \eqref{Analytical F}.

We have also provided expressions for the perturbed fluid variables in Generalized Einstein-Aether models, in the scalar, vector, and tensor sectors. These vector-tensor theories have non-canonical kinetic terms and are modified by a free function, $\mathcal{F}(\mathcal{K})$. While some work has been done on these theories, the $c_4$ term in \eqref{Kinetic Tensor} is often set to zero. It is often argued that this can be done via a redefinition of the coefficients, which is true only if the Aether field is hypersurface orthogonal i.e. as in the Khronometric model \eqref{Khronon}. A consequence of this is that no transverse vector mode propagates at the level of linear perturbations. To keep things more general we keep the $c_4$ term in our analysis.

The EoS approach to cosmological perturbations provides a way of parametrizing dark energy models and modified gravity theories via the gauge invariant entropy perturbation and anisotropic stresses. This is done by fully eliminating the internal degrees of freedom introduced by this theory. In this paper, we have provided expressions for these in terms of linear functions of the perturbed variables and metric perturbations, $\Pi^S_{\mathrm{de}}=\Pi^S_{\mathrm{de}}(\Delta_{\mathrm{de}},\hat{\Theta}_{\mathrm{de}},X,Y)$ and $\Gamma_{\mathrm{de}}=\Gamma_{\mathrm{de}}(\Delta_{\mathrm{de}},\hat{\Theta}_{\mathrm{de}},W,X,Y)$, given in \eqref{Pi} and \eqref{Gamma}. They have been expressed in an explicitly gauge invariant form thanks to a new set of notation. Furthermore, via the Einstein equations, we are also able to specify them in terms of the perturbed fluid variables for dark energy and matter only i.e. $\Pi^S_{\mathrm{de}}=\Pi^S_{\mathrm{de}}(\Delta_{\mathrm{de}},\Delta_{\mathrm{m}},\hat{\Theta}_{\mathrm{de}},\hat{\Theta}_{\mathrm{m}},\Pi_{\mathrm{m}})$ and $\Gamma_{\mathrm{de}}=\Gamma_{\mathrm{de}}(\Delta_{\mathrm{de}},\Delta_{\mathrm{m}},\hat{\Theta}_{\mathrm{de}},\hat{\Theta}_{\mathrm{m}},\Gamma_{\mathrm{m}})$, given by \eqref{Pi DM} and \eqref{Gamma DM}. We note that there seems to be a discontinuity in taking the $\Lambda$CDM limit in a designer $\mathcal{F}(\mathcal{K})$ model. From these, we solve for the evolution of the Newtonian gravitational potentials via the perturbed fluid equations for varying parameters, shown in \autoref{fig:ZYk} and \autoref{fig:Phi Potential a}. In a designer $\mathcal{F}(\mathcal{K})$ we find that $w_\mathrm{de} \Pi_{\mathrm{de}}^S \rightarrow 0$ for $K \gg 1 $, independent of the choice of $\left\lbrace c_i \right\rbrace$. We also provide expressions for $\Pi^{V,T}$ in the vector and tensor sectors, given by \eqref{Pi V} and \eqref{Pi T}. 

Of course, the motivation for this analysis is to obtain observables in cosmology and see how they compare to $\Lambda\mathrm{CDM}$. We have now provided the necessary expressions in order to solve the perturbed fluid equations and obtain spectra. In principle, this should be easy to incorporate into existing numerical codes. Similar to \cite{EOS1,EOS2,EOS3}, we would like to explore a broader class of vector-tensor models, without ever having to specify a specific model. What if we know nothing about the background Lagrangian other than its field content? Can anything be said more broadly about general vector-tensor theories of gravity and their application to dark energy? This is similar to work done in \cite{GeneralSVT}, but instead adopting a covariant approach as was done in \cite{EOS3} for scalar-tensor theories. We leave this as a matter for future work.

\section*{Acknowledgements}
We would like to thank Boris Bolliet for very helpful discussions and comments. DT is supported by an STFC studentship. FP is supported by an STFC postdoctoral fellowship.

\appendix
\section{Equations of state for perturbations in the synchronous gauge} \label{app:A}

In the synchronous gauge, we have that \begin{align}
a^2 \delta \rho =& \mathcal{F}_\mathcal{K}\left[ \hspace{1mm} c_{14}k^2 V'  +\left( c_{14}-\alpha(1+2\gamma_2)\right)\mathcal{H}k^2V+\frac{1}{2}\alpha\mathcal{H}(1+2\gamma_2)h'\right] ,\\ 
a^2\rho(1+w_\mathrm{de}) \theta^S =& \frac{1}{6}\mathcal{F}_\mathcal{K}\left[ (2k^2V-h')(3c_{123}+2\alpha \gamma_2)-12c_{13}\eta'\right], 
\end{align} where \begin{equation}
\delta \mathcal{K} = \frac{2 \alpha\mathcal{H}}{a^2 M^2}\left( \frac{1}{2}h'-k^2 V\right).
\end{equation} 

We can then write this system of equations as \begin{equation} 
a^2\begin{pmatrix}
\delta\rho\\
\rho (1+w_\mathrm{de})\theta^S
\end{pmatrix} = k^2\begin{pmatrix}
A & B \\
0 & C
\end{pmatrix} \begin{pmatrix}
V'\\
V
\end{pmatrix} + \begin{pmatrix}
D\\
E
\end{pmatrix},
\end{equation}  with \begin{align}
A &= c_{14}\mathcal{F}_\mathcal{K}, \\
B &=\left[ c_{14}-\alpha(1+2\gamma_2)\right] \mathcal{H}\mathcal{F}_\mathcal{K},\\
C &=\frac{1}{3}\mathcal{F}_\mathcal{K}(3c_{123}+2\alpha \gamma_2), \\
D &= \frac{1}{2}\alpha \mathcal{H}\mathcal{F}_\mathcal{K}(1+2\gamma_2)h', \\
E &= -\frac{1}{6}\mathcal{F}_\mathcal{K}\left[ (3c_{123}+2\alpha \gamma_2)h'+12c_{13}\eta' \right]. 
\end{align} Inverting this will give us expressions for $V$ and $V'$ in terms of $\delta \rho$, $\theta^S$, the metric perturbations, $h$ and $\eta$, and their derivatives. Eliminating for these in $\Pi^S$ \eqref{Sync Pi}, we find that we can write \eqref{general Pi} as \begin{equation}
w\Pi^S = c_{\Pi \Delta} \Delta + c_{\Pi \Theta}\hat{\Theta} + c_{\Pi X}X + c_{\Pi Y}K^2 Y,
\end{equation} where the $c_\Pi$ coefficients are given in \eqref{c_PiD} to \eqref{c_PiY}. In order to show this, we use the conservation equation \eqref{Background ConsEq} to find that \begin{equation} \label{Background ConsEq 2}
3(1+w_\mathrm{de})=\epsilon_H\frac{2\gamma_1(1+2\gamma_2)}{2\gamma_1-1}
\end{equation} and replace for this in $3(1+w)T$, arising from $\hat{\Theta}$ in \autoref{table:GInv Var}. From this it can be shown that the coefficient $c_{\Pi T}=0$, as discussed previously.

Similarly, we do the same for the entropy perturbation by eliminating $V$ and $V'$ in $\delta P$ and hence find that \begin{equation}
w\Gamma = c_{\Gamma\Delta} \Delta + c_{\Gamma\Theta}\hat{\Theta} + c_{\Gamma W} W + c_{\Gamma X} X + c_{\Gamma Y} K^2 Y,
\end{equation} where the $c_\Gamma$ coefficients are as before in \eqref{c_GammaD} to \eqref{c_GammaY}. To show this, we note that there is a term proportional to $\displaystyle3(1+w_\mathrm{de})\frac{dP}{d\rho}T$. As before, we use \eqref{Background ConsEq 2} to replace $3(1+w_\mathrm{de})$ and also compute that \begin{equation}
\frac{dP}{d\rho}=\frac{a^2P'}{a^2\rho'}=\epsilon_H\left(  \frac{2\gamma_2}{1+2\gamma_2}\right) \left( 1+\frac{2}{3}\gamma_3\right) +\frac{2}{3}\epsilon_H -1   -\frac{\epsilon_H\hspace{0.1mm}'}{3\mathcal{H}\epsilon_H}.
\end{equation} After substituting in for these it can be shown that $c_{\Gamma T}=0$. Hence, \eqref{Pi} and \eqref{Gamma} constitute the gauge invariant equations of state for the perturbations.

\section{Constraints on coefficients in Minkowski space} \label{app:B}

\begin{table}
	\begin{center}
		{\renewcommand{\arraystretch}{1.6}\begin{tabular}{l | c}
				\multicolumn{1}{c|}{\textbf{Constraints}}& \textbf{ Reason} \\
				\hline
				(a)$\hspace{2mm}0\leq\dfrac{c_{123}}{c_{14}} \leq 1$& Non-tachyonic and subluminal propagation of scalar modes \\
				(b)$\hspace{2mm}0\leq\dfrac{c_1}{c_{14}} \leq 1$& Non-tachyonic and subluminal propagation of vector modes \\
				(c)$\hspace{2mm}c_{13} \geq 0$& Subluminal propagation of gravitational waves\\
				(d)$\hspace{2mm}c_{14}< 0$& No ghosts\\
				(e)$\hspace{2mm}c_{123}\leq0$& (a) and (d) \\
				(f)$\hspace{2mm}c_1\leq0$ and $c_4 \geq0$& (b) and (d) \\
				(g)$\hspace{2mm}c_2 \leq0$& (c) and (e) \\
				(h)$\hspace{2mm}c_3 \geq0$&(c) and (f) \\
				(i)$\hspace{2mm}\alpha \leq 0$&(e) and (g)
		\end{tabular}}
	\end{center}
	\caption{Summary of the constraints on the $\left\lbrace c_i\right\rbrace $ coefficients, obtained from Minkowski space and gravitational waves.} \label{table:constraints1}
\end{table}

We would like to obtain constraints on the $\left\lbrace c_i\right\rbrace $ coefficients by studying the behaviour of perturbations in Minkowski space. We largely follow the procedure defined in \cite{Lim}, extending their results to include $c_4\not=0$. The Lagrangian which governs the perturbations is obtained by perturbing the degrees of freedom in the background Lagrangian to quadratic order. This would then give rise to linear equations of motion for the perturbations. Schematically, we are computing $\mathcal{L}\rightarrow\mathcal{L}+\delta\mathcal{L}+\delta^2\mathcal{L}$, where $\delta^2 \mathcal{L}$ denotes the Lagrangian quadratic in perturbations. Again suppressing over-bars to denote unperturbed variables, from \eqref{General Lagrangian} we have that \begin{equation}
\delta^2 \mathcal{L} = M^2 \left( \mathcal{F}_{\mathcal{KK}}(\delta \mathcal{K})^2 + \mathcal{F}_\mathcal{K}\delta^2\mathcal{K}\right) + 2 A^\mu \delta A_\mu \delta \lambda,
\end{equation} since $\lambda = 0$ in Minkowski space.

Perturbing the Aether as $A^\mu \rightarrow A^\mu + \delta A^\mu = (1,0,0,0)+v^\mu$ and assuming the metric to be flat, we can compute $M^2\delta^2 \mathcal{K}$ by perturbing the Aether field and expanding out to quadratic order, to give \begin{equation}
M^2 \delta^2 \mathcal{K} = c_1 \partial_\mu v^\nu \partial^\mu v_\nu + c_2 (\partial_\mu v^\mu)^2 +c_3 \partial_\mu v^\nu \partial_\nu v^\mu + c_4 A^\mu A^\nu \partial_\mu v^\rho \partial_\nu v_\rho + 2\delta\lambda A^\mu v_\mu.
\end{equation} Similarly we can calculate $M^2 \delta K$ to be \begin{equation}
\frac{1}{2}M^2\delta \mathcal{K} = c_1 \partial_\mu A^\nu \partial^\mu v_\nu + c_2 \partial_\mu A^\mu \partial_\nu v^\nu + c_3 \partial_\mu A^\nu \partial_\nu v^\mu +c_4 A^\mu v^\nu \partial_\mu A^\rho \partial_\nu A_\rho + c_4 A^\mu A^\nu \partial_\mu A^\rho \partial_\nu v_\rho.
\end{equation} From this we see that in Minkowski space $\delta \mathcal{K} = 0$ since $\partial_\mu A^\nu =0$, which will also be true for the unperturbed value of $\mathcal{K}$. The second order Lagrangian is therefore given by \begin{equation} \label{SecondOrderL1}
\delta^2 \mathcal{L} = \mathcal{F}_\mathcal{K}\left[ -c_{14} \dot{v}^2 + c_1 \partial_i v^j \partial^i v_j + c_2 (\partial_i v^i)^2 +c_3 \partial_i v^j \partial_j v^i\right] ,
\end{equation} where $\dot{v}^2 = \dot{v}^i \dot{v}_i$ and we have used $v^0 = 0$. By analogy to the cosmological perturbations, we decompose the perturbation into a scalar and vector part, \begin{equation} \label{decomp1}
v^i = \partial^i V + iB^i = S^i + T^i,
\end{equation} such that $k^i T_i = 0$. Inserting this into \eqref{SecondOrderL1}, we find that we can write it as the sum of two uncoupled Lagrangians for the fields $S^i$ and $T^i$, since any cross terms are zero by the scalar-vector decomposition of the perturbation. They are given by
\begin{equation} \label{Spin-0 L1}
\mathcal{L}_S = \mathcal{F}_\mathcal{K} \left[ -c_{14} \dot{S}^2 + c_1 \partial_i S^j \partial^i S_j + c_2 (\partial_i S^i)^2 +c_3 \partial_i S^j \partial_j S^i \right],
\end{equation}   \begin{equation} \label{Spin-1 L1}
\mathcal{L}_T = \mathcal{F}_\mathcal{K} \left[ -c_{14}\dot{T}^2 + c_1 \partial_i T^j \partial^i T_j \right].
\end{equation} Here we see the problem with the Minkowski limit for the the designer model, with $\mathcal{F} = B(\mathcal{K})^{1/2}+C$. Since $\mathcal{K}\propto H^2$, in the Minkowski limit where $\mathcal{K} \rightarrow 0$ we have that $\mathcal{F}_\mathcal{K} \rightarrow \infty$ and hence the second order Lagrangian is not well defined. Constraints can still be obtained for the $\left\lbrace c_i\right\rbrace $ coefficients, but not for the designer model. To compare with results from \cite{Lim,JacobsonAEWaves} we will set $\mathcal{F}_\mathcal{K} = 1$.

Hence, the equations of motion from \eqref{Spin-0 L1} and \eqref{Spin-1 L1} are then given by \begin{equation} \label{Wave S1}
\ddot{S}_i - \frac{c_{123}}{c_{14}}\partial^j \partial_j S_i = 0, \quad \ddot{T}_i - \frac{c_1}{c_{14}}\partial^j \partial_j T_i = 0,
\end{equation}  where we have used $\partial_i S_j = \partial_j S_i$ from the definition in \eqref{decomp1}. Therefore, we see that $S_i$ and $T_i$ propagate with sound speeds 
$\displaystyle c^2_s = \frac{c_{123}}{c_{14}}$ and $\displaystyle c_s ^2 = \frac{c_1}{c_{14}}$ respectively. Imposing that the propagation speeds are less than $c$ and to avoid them being imaginary, leading to an exponential growth in perturbations, we require \begin{equation}
0 \leq \frac{c_{123}}{c_{14}} \leq 1 \hspace{5mm} \mathrm{and} \hspace{5mm} 0 \leq \frac{c_1}{c_{14}} \leq 1
\end{equation} 

Also, following the process of \cite{Lim}, considerations of the quantum Hamiltonian gives an additional constraint of $c_{14} < 0$ to prevent ghosts. Heuristically we can see this from \eqref{SecondOrderL1}, as $c_{14} < 0$ ensures that the kinetic term is the correct sign, however see \cite{Lim} for a full treatment of the quantization of this theory.
%The solution to \eqref{Wave S} is simply a plane wave, \begin{equation} %\label{Wave sol}
%S_i \propto \exp \left[ -i(c_s kt - k_i x^i) \right].
%\end{equation}

%We see in \eqref{Spin-0 L} and \eqref{Spin-1 L}, to have the correct sign for the kinetic term in the Lagrangian we need $(c_4 - c_1) > 0$. The Hamiltonian for the field $S_i$ is $\mathcal{H}_S = \Pi^i_{S} \dot{S}_i - \mathcal{L}_S$, where $\Pi^i _S = \dfrac{\partial \mathcal{L}}{\partial \dot{S}_i}$ is the conjugate momentum and similarly for $T_i$. We find that $\Pi_i = 2(c_4 - c_1) \dot{S}_i$ and hence \begin{equation}
%\mathcal{H}_S = -\left[ (c_1-c_4) \dot{S}^2 + c_1 \partial_i S^j \partial^i S_j + c_2 (\partial_i S^i)^2 +c_3 \partial_i S^j \partial_j S^i \right] .
%\end{equation} Inserting the solution \eqref{Wave sol} we find that \begin{equation}
%\mathcal{H}_S = 2(c_1-c_4)c_s^2k^2S^2.
%\end{equation} Recalling that $S^i = \partial^i V$, or in Fourier space $S^i = ik^i V$, then \begin{equation}
%\mathcal{H}_S = 2(c_4 - c_1)c_s^2k^4V^2.
%\end{equation} Therefore, for the Hamiltonian to be positive definite, we require \begin{equation}
%(c_4 - c_1) > 0.
%\end{equation} Similarly, for the field $B_i$ we obtain \begin{equation}
%\mathcal{H}_T =  -\left[ (c_1-c_4) \dot{T}^2 + c_1 \partial_i T^j \partial^i T_j \right]
%\end{equation} and again inserting a plane wave solution, \begin{equation}
%\mathcal{H}_T = 2(c_1-c_4)k^2T^2 = 2(c_4-c_1)k^2B^2
%\end{equation} and so again we conclude that we require \begin{equation}
%(c_4 - c_1) > 0.
%\end{equation} 

Let us summarise the constraints we have obtained. As in \cite{Lim}, we can also infer further constraints from those already obtained, allowing us to get more useful constraints on the individual coefficients and also combinations of them that appear frequently. They are are shown in \autoref{table:constraints1} and are also consistent with those obtained in \cite{JacobsonAEWaves}.

\label{lastpage}

\bibliographystyle{apsrev4-1}
%\bibliography{../../Bibliography/old_MasterBib}
\bibliography{GEA_final.bbl}
\end{document}